\documentclass[11pt,a4paper]{article}
\usepackage[utf8]{inputenc}
\usepackage[english]{babel}
\usepackage[top=2.5cm, bottom=3.0cm, left=2.5cm, right=2.5cm]{geometry}
\usepackage{setspace}
\usepackage{dutchcal}
\makeatletter
\g@addto@macro\bfseries{\boldmath}
\makeatother
\usepackage{fancyhdr}
\usepackage{amssymb}
\usepackage{amsmath}
\usepackage{amsfonts}
\usepackage{mathrsfs}
\usepackage{bm}
\usepackage{slashed}
\usepackage{mathtools}
\numberwithin{equation}{section}
\allowdisplaybreaks
\usepackage{enumitem}
\usepackage{array}

\usepackage[]{caption} 
\usepackage{hyperref}
\hypersetup{
    colorlinks,%
    citecolor=blue,%
    filecolor=blue,%
    linkcolor=blue,%
   	urlcolor=blue,
   	linktoc=page
}
\usepackage{cite}
\usepackage[toc,page]{appendix}
\usepackage{graphicx}
\usepackage{float}
\usepackage{subfig}
\usepackage{tikz}
\usetikzlibrary{calc}
\usetikzlibrary{decorations.pathreplacing}
\usetikzlibrary{decorations.pathmorphing}
\usetikzlibrary{decorations.markings}
\usetikzlibrary{arrows.meta}
\tikzset{every picture/.style={font issue=\footnotesize},
         font issue/.style={execute at begin picture={#1\selectfont}}
        }
\usepackage{xcolor}
\definecolor{DarkGreen}{RGB}{0,128,0}
\usepackage{mdframed}
\newcommand{\p}{\partial}

\newcommand{\ndelta}{\delta\hspace{-0.50em}\slash\hspace{-0.05em} }
\newcommand{\poubelle}[1]{}

\renewcommand{\textbf}[1]{\begingroup\bfseries\mathversion{bold}#1\endgroup}
\newcommand{\D}{\text d}

\usepackage{pict2e}
\makeatletter
\newcommand{\loplus}{\mathbin{\mathpalette\dog@lsemi{+}}}
\newcommand{\dog@lsemi}[2]{\dog@semi{#1}{#2}{270,90}}
\newcommand{\dog@semi}[3]{%
  \begingroup
  \sbox\z@{$\m@th#1#2$}%
  \setlength{\unitlength}{\dimexpr\ht\z@+\dp\z@\relax}%
  \makebox[\wd\z@]{\raisebox{-\dp\z@}{%
    \begin{picture}(1,1)
    \linethickness{\variable@rule{#1}}
    \roundcap
    \put(0.5,0.5){\makebox(0,0){\raisebox{\dp\z@}{$\m@th#1#2$}}}
    \put(0.5,0.5){\arc[#3]{0.5}}
    \end{picture}%
  }}%
  \endgroup
}
\newcommand{\variable@rule}[1]{%
  \fontdimen8  
  \ifx#1\displaystyle\textfont3\else
    \ifx#1\textstyle\textfont3\else
      \ifx#1\scriptstyle\scriptfont3\else
        \scriptscriptfont3\relax
  \fi\fi\fi
}
\makeatother

\usepackage{ulem}

\begin{document}

\setstretch{1.1}

\setcounter{tocdepth}{2}

\begin{titlepage}

\begin{flushright}\vspace{-3cm}
{\small
\phantom{\today} }\end{flushright}

\begin{center}
\setstretch{1.75}
{\LARGE{\textbf{ 
Charge Algebra in Al(A)dS$_\text{\itshape n}$ Spacetimes
}}}
\end{center}
 \vspace{7mm}
\begin{center} 
\centerline{\large{\bf{Adrien Fiorucci$^\star$\footnote{e-mail: adrien.fiorucci@ulb.be}, Romain Ruzziconi$^\dagger$\footnote{e-mail: romain.ruzziconi@tuwien.ac.at}}}}

\vspace{2mm}
\normalsize
\bigskip\medskip
$^\star$\textit{Universit\'{e} Libre de Bruxelles and International Solvay Institutes\\
CP 231, B-1050 Brussels, Belgium\\
\vspace{2mm}
}

\vspace{2mm}
\normalsize
\bigskip\medskip
$^\dagger$\textit{Institute for Theoretical Physics, TU Wien,\\
Wiedner Hauptstrasse 8, A-1040 Vienna, Austria\\
\vspace{2mm}
}

\vspace{10mm}

\begin{abstract}
The gravitational charge algebra of generic asymptotically locally (A)dS spacetimes is derived in $n$ dimensions. The analysis is performed in the Starobinsky/Fefferman-Graham gauge, without assuming any further boundary condition than the minimal falloffs for conformal compactification. In particular, the boundary structure is allowed to fluctuate and plays the role of source yielding some symplectic flux at the boundary. Using the holographic renormalization procedure, the divergences are removed from the symplectic structure, which leads to finite expressions. The charges associated with boundary diffeomorphisms are generically non-vanishing, non-integrable and not conserved, while those associated with boundary Weyl rescalings are non-vanishing only in odd dimensions due to the presence of Weyl anomalies in the dual theory. The charge algebra exhibits a field-dependent $2$-cocycle in odd dimensions. When the general framework is restricted to three-dimensional asymptotically AdS spacetimes with Dirichlet boundary conditions, the $2$-cocycle reduces to the Brown-Henneaux central extension. The analysis is also specified to leaky boundary conditions in asymptotically locally (A)dS spacetimes that lead to the $\Lambda$-BMS asymptotic symmetry group. In the flat limit, the latter contracts into the BMS group in $n$ dimensions.
\\[1.5cm]

\noindent \textit{Keywords}: Asymptotic symmetries, charge algebra, Starobinsky/Fefferman-Graham gauge, asymptotically locally (A)dS spacetimes, leaky boundary conditions, Weyl charges, conformal anomalies, central extension, holography with sources, BMS group, covariant phase space, holographic renormalization.

\end{abstract}



\end{center}

\end{titlepage}

\newpage

\setcounter{page}{2}

\begin{spacing}{0.8}

\tableofcontents

\end{spacing}

\newpage

\section{Introduction}
\label{sec:intro}

Asymptotically anti-de Sitter (AdS) spacetimes have been the focus of constant attention these last decades due to their crucial role in the AdS/CFT correspondence. The study of asymptotic symmetries in the gravity side of the duality has been shown to be extremely efficient to extract some patterns of the holographic correspondence. One of the pioneer works in this direction was the investigation of Dirichlet boundary conditions in three dimensions, where the asymptotic symmetry algebra is the infinite-dimensional conformal algebra in two dimensions \cite{Brown:1986nw}. The associated charge algebra was shown to be a double copy of the Virasoro algebra with the famous Brown-Henneaux central extension. This central extension was then related to the Weyl anomaly of the dual CFT \cite{Henningson:1998gx}. Furthermore, it was used to reproduce the BTZ black hole entropy using the Cardy formula \cite{Strominger:1997eq}. The Dirichlet boundary conditions were also considered in higher dimensions, where the conformal algebra always appears \cite{Hawking:1983mx , Ashtekar:1984zz , Henneaux:1985tv , Henneaux:1985ey , Ashtekar:1999jx} (see also \cite{Fischetti:2012rd} and references therein). Other boundary conditions in asymptotically AdS spacetimes were proposed in the literature \cite{Troessaert:2013fma , Compere:2013bya , Papadimitriou:2005ii , Grumiller:2016pqb , Perez:2016vqo , Aros:1999kt , Alessio:2020ioh}, leading to different asymptotic symmetries and holographic dualities.

Motivated by various considerations, a set of boundary conditions was for a long time seen as reasonable if it led to an action that is stationary on solutions and charges that are integrable and conserved in time. However, some developments suggest us to drop out these strong requirements. A first example occurs in the recent analysis of the black hole information paradox to derive the Page curve from quantum gravity path integral arguments \cite{Almheiri:2019yqk , Almheiri:2019qdq , Almheiri:2020cfm}. In this context, it has been useful to allow some radiation to escape the spacetime boundary so that the black hole can evaporate in AdS. This was implemented in practice by gluing an asymptotically flat region to the AdS boundary and coupling the dual theory to a thermal bath (see Figure \hyperlink{fig:AdS}{1.(a)}). Another example appears when considering brane worlds interacting with ambient higher-dimensional spacetimes \cite{Randall:1999ee, Randall:1999vf}. This picture naturally yields holographic dualities with fluctuating boundary metric and induced quantum gravity on the boundary \cite{Compere:2008us}. 

These examples are appealing to investigate a more general class of leaky boundary conditions in AdS that allow for some flux at infinity and fluctuating boundary structure. Technical results have already been developed to investigate non-integrable and non-conserved charges in the context of asymptotically flat spacetimes \cite{Barnich:2011mi , Barnich:2013axa}, where considering a non-vanishing flux at infinity is mandatory to treat the gravitational radiation \cite{Sachs:1962wk , 1962RSPSA.269...21B , Penrose:1965am , Newman:1968uj , Ashtekar:2014zsa}. In particular, a modified bracket to deal with non-integrable charge expressions has been proposed in order to compute the charge algebra, and then successfully applied in several different contexts, including in asymptotically locally flat spacetimes \cite{Compere:2018ylh , Barnich:2019vzx} and in the study of asymptotic symmetries at the black hole horizon \cite{Donnay:2016ejv , Adami:2020amw , Chandrasekaran:2020wwn}. One of the purposes of this work is to import these techniques into asymptotically locally AdS spacetimes and generalize part of the analysis performed in \cite{Compere:2020lrt} to arbitrary spacetime dimensions.  

While the holographic correspondence is less clear in asymptotically de-Sitter (dS) spacetimes \cite{Strominger:2001pn , Anninos:2011ui, Park:1998qk}, the latter are of major interest in cosmology since they offer simple mathematical models for the observed accelerating expansion of the universe. Imposing boundary conditions at the future spacelike boundary $\mathscr{I}_{\text{dS}}^+$  \cite{Anninos:2011jp , Anninos:2010zf , Ashtekar:2019khv , Ashtekar:2014zfa , Compere:2019bua} (see Figure \hyperlink{fig:dS}{1.(b)}) is a delicate issue since it can drastically restrict the Cauchy problem by eliminating late-time radiation. One is naturally led to allow some non-vanishing flux going through the spacetime boundary and leaky boundary conditions become an essential ingredient. Therefore, we will cover asymptotically locally dS spacetimes as well in our general analysis. While the mathematical formulae are essentially the same than in the asymptotically locally AdS case, the interpretation of the results is expected to be completely different. 

\begin{figure}[h!]
\centering
\begin{tikzpicture}[scale=0.8]
\draw (0,8)node[]{\hypertarget{fig:AdS}{}};
\draw (8,3)node[]{\hypertarget{fig:dS}{}};
\draw (0,-8) node[below]{\normalsize (a) \ AlAdS \textit{case}.};
\draw (8,-8) node[below]{\normalsize (b) \ AldS \textit{case}.};


\draw[white] (-4,-9) -- (-4,8) -- (12,8) -- (12,-9) -- cycle;
 
\def\dx{.2};
\def\dy{.4};
\coordinate (A) at (-2,-7+\dy);
\coordinate (B) at (2,-7+\dy);
\coordinate (C) at (2,-\dy);
\coordinate (D) at (-2,-\dy);

\coordinate (E) at (-2,\dy);
\coordinate (F) at (2,\dy);
\coordinate (G) at (2,7-\dy);
\coordinate (H) at (-2,7-\dy);

\draw (B) arc[x radius=2, y radius=0.5, start angle=0, end angle=-180];
\draw [dashed] (B) arc[x radius=2, y radius=0.5, start angle=0, end angle=180];
\draw (A) -- (D);
\draw (B) -- (C);
\draw (C) arc[x radius=2, y radius=0.5, start angle=0, end angle=-360];

\draw (E) -- (H);
\draw (F) -- (G);
\fill [white] (F) arc[x radius=2, y radius=0.5, start angle=0, end angle=360];
\draw (F) arc[x radius=2, y radius=0.5, start angle=0, end angle=-180];
\draw [dashed] (F) arc[x radius=2, y radius=0.5, start angle=0, end angle=180];
\draw (G) arc[x radius=2, y radius=0.5, start angle=0, end angle=360];

\draw [dashed] (D) -- (E);
\draw [dashed] (C) -- (F);
\draw [dashed] (A) -- ($(A)-(0,2*\dy)$);
\draw [dashed] (B) -- ($(B)-(0,2*\dy)$);
\draw [dashed] (G) -- ($(G)+(0,2*\dy)$);
\draw [dashed] (H) -- ($(H)+(0,2*\dy)$);

\def\decal{0.1};
\draw [decorate,decoration={brace,amplitude=10pt}] ($(E)-(\decal,0)$) -- ($(H)-(\decal,0)$) node[midway,xshift=-25,draw=black,align=center,rotate=90,text width=2.3cm,fill=black!5] {\textbf{Leaky b.c.}}; 
\draw [decorate,decoration={brace,amplitude=10pt}] ($(A)-(\decal,0)$) -- ($(D)-(\decal,0)$) node[midway,xshift=-25,draw=black,align=center,rotate=90,text width=3.5cm,fill=black!5] {\textbf{Conservative b.c.}}; 

\path [-{Latex[width=2mm]},thick,draw=red] (0.5,-5.5) -- (2,-4) -- (0,-2);
\path [-{Latex[width=2mm]},thick,draw=red] (0.25,-5.25) -- (2,-3.5) -- (0.25,-1.75);
\path [-{Latex[width=2mm]},thick,draw=red] (0,-5) -- (2,-3) -- (0.5,-1.5);
\path [blue,thick] (-2,-4.5) edge[bend left=30] (0,-4.5) -- (0,-4.5) edge[bend right=30] (2,-4.5);
\draw (-1,-4) node[above,blue]{$\Sigma$};

\path [-{Latex[width=2mm]},thick,draw=red] (0.5,1.5) -- (2,3) -- (3.25,4.25);
\path [-{Latex[width=2mm]},thick,draw=red] (0.25,1.75) -- (2,3.5) -- (3,4.5);
\path [-{Latex[width=2mm]},thick,draw=red] (0,2) -- (2,4) -- (2.75,4.75);
\path [blue,thick] (-2,2.5) edge[bend left=30] (0,2.5) -- (0,2.5) edge[bend right=30] (2,2.5);
\draw (-1,3) node[above,blue]{$\Sigma$};

\coordinate (C1) at ($(F)!0.9!(G)$);
\coordinate (C2) at ($(F)!0.5!(G)$);
\coordinate (C3) at ($(F)!0.1!(G)$);
\coordinate (D1) at ($(C1)+(3*\decal,0)$);
\coordinate (D2) at ($(C2)+(2.5+3*\decal,0)$);
\coordinate (D3) at ($(C3)+(3*\decal,0)$);

\draw[black!50,very thick] (D1) -- node[black,above,midway]{$\qquad\mathscr I^+$} (D2)node[right,black]{$i^0$} -- node[black,below,midway]{$\qquad\mathscr I^-$} (D3);

\draw ($(D2)!0.5!(D3)$) node[rotate=45,anchor=south]{\textbf{Flat spacetime}};

\draw ($(A)!0.5!(B)$) node[]{$i^-_{\text{AdS}}$};
\draw ($(G)!0.5!(H)$) node[]{$i^+_{\text{AdS}}$};
\draw ($(B)!0.8!(C)$) node[right]{$\mathscr I_{\text{AdS}}$};



\coordinate (P) at (5,-7);
\coordinate (Q) at (11,-7);
\coordinate (R) at (11,-1);
\coordinate (S) at (5,-1);
\draw (P) -- (Q) -- (R) -- (S) -- cycle;
\draw ($(P)!0.8!(Q)$) node[below]{$\mathscr I^-_{\text{dS}}$};
\draw ($(R)!0.8!(S)$) node[above]{$\mathscr I^+_{\text{dS}}$};
\draw ($(P)!0.5!(S)$) node[above,rotate=90]{North pole};
\draw ($(Q)!0.5!(R)$) node[above,rotate=-90]{South pole};
\path [blue,thick] (5,-3) edge[bend left=30] (8,-3) -- (8,-3) edge[bend right=30] (11,-3);
\draw (9.5,-4) node[blue]{$\Sigma$};
\draw[black!50,dashed] (P) -- (R);

\path [-{Latex[width=2mm]},thick,draw=red] (6,-3.5) -- (10,0.5);
\path [-{Latex[width=2mm]},thick,draw=red] (6.25,-3.75) -- (10.25,0.25);
\path [-{Latex[width=2mm]},thick,draw=red] (6.5,-4) -- (10.5,0);

\draw (8,1.5) node[draw=black,align=center,text width=2.3cm,fill=black!5] {\textbf{Leaky b.c.}};
\draw [->] (8,1) -- (8,0);

\end{tikzpicture}
\caption{Conservative \textit{vs.} leaky boundary conditions in Al(A)dS spacetimes.}
\label{fig:Conservative vs leaky}
\end{figure}
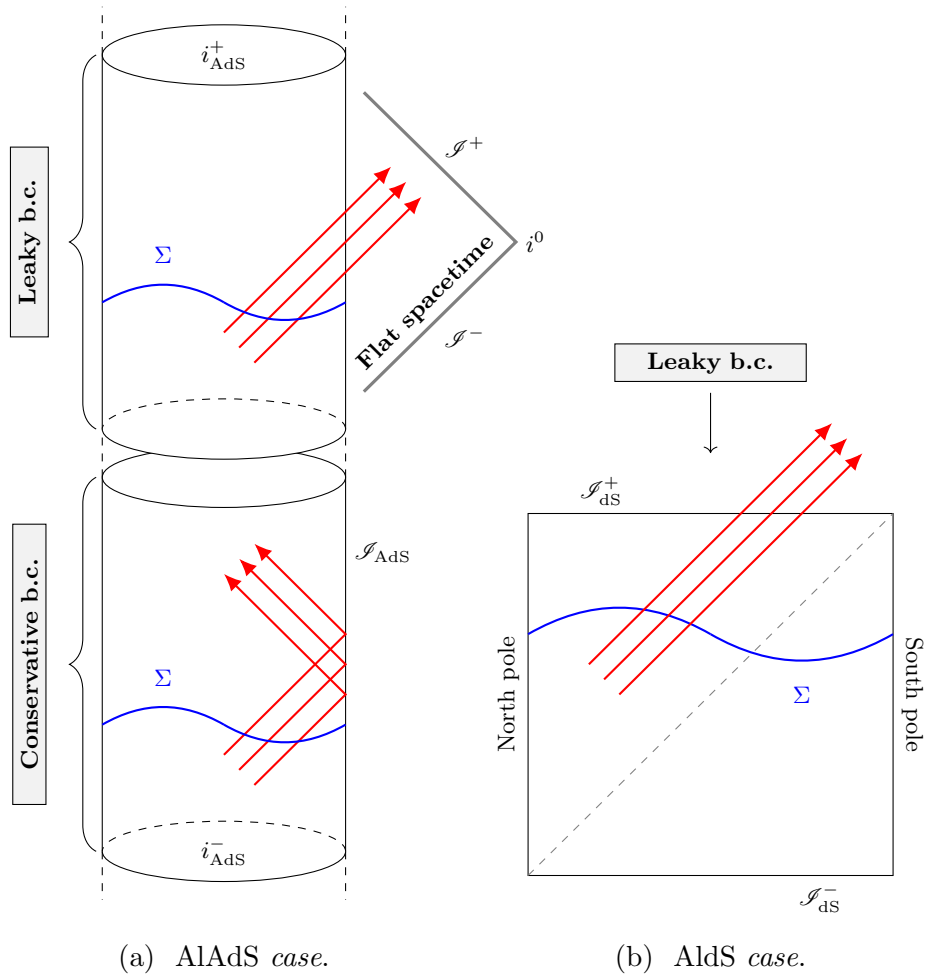


In this work, we study the phase space of the most generic asymptotically locally (A)dS (Al(A)dS) spacetimes in $n = d+1$ dimensions ($n \ge 3$). We work in the Starobinsky/Fefferman-Graham gauge \cite{Starobinsky:1982mr,Fefferman:1985aa}, which is particularly well adapted for the study of these spacetimes and where the holographic dictionary is well established \cite{deHaro:2000vlm , Skenderis:2002wp}. We assume the minimal falloffs conditions that allow for conformal compactification and a clear definition of what the asymptotic region means. After reviewing results on the solution space and its transformation under residual gauge diffeomorphisms, we use the holographic renormalization procedure \cite{deHaro:2000vlm , Skenderis:2002wp} to remove the divergences in the expansion parameter appearing in the symplectic structure \cite{Compere:2008us}. From the renormalized symplectic form, we derive the charges of the theory. The boundary diffeomorphism charges are generically non-vanishing and non-integrable in any spacetime dimension. By contrast, the Weyl charges vanish in even dimensions and are generically non zero in odd dimensions (their explicit expressions are derived explicitly in three and five spacetime dimensions). We interpret this fact as related to the presence of Weyl anomalies in the dual theory \cite{Henningson:1998gx}. Indeed, Weyl anomalies prevent from choosing freely any representative of the boundary metric in the conformal class obtained by conformal compactification. In the bulk, this translates into the fact that residual gauge diffeomorphisms inducing Weyl rescalings on the boundary are large gauge transformations in odd dimensions. Then we compute the charge algebra using the Barnich-Troessaert bracket \cite{Barnich:2011mi}. Interestingly, similarly to the four-dimensional asymptotically flat case, the charge algebra exhibits a field-dependent $2$-cocycle in odd spacetime dimensions. When imposing more restrictive Dirichlet boundary conditions, we show that this $2$-cocycle reduces to the Brown-Henneaux central extension in three dimensions \cite{Brown:1986nw}. We also apply our general analysis to discuss leaky boundary conditions in asymptotically (A)dS spacetimes that yield the analogue of the $\Lambda$-BMS$_4$ asymptotic symmetry algebra discussed in \cite{Compere:2019bua , Compere:2020lrt} in $n$ dimensions, written $\Lambda$-BMS$_n$. This $\Lambda$-BMS$_n$ algebra has the interesting property to reduce to the (generalized) BMS$_n$ algebra in the flat limit \cite{1962RSPSA.269...21B , Sachs:1962wk , Barnich:2010eb , Campiglia:2014yka , Campiglia:2015yka , Compere:2018ylh , Colferai:2020rte , Kapec:2015vwa, Campoleoni:2017qot , Aggarwal:2018ilg , Barnich:2006av , Hollands:2016oma,Flanagan:2015pxa,Flanagan:2019vbl}. Furthermore, the $\Lambda$-BMS$_n$ algebra may be deeply involved in the investigation of the infrared structure of gravity in presence of non-vanishing cosmological constant \cite{Hinterbichler:2013dpa,Berezhiani:2013ewa,Horn:2014rta,Mirbabayi:2016xvc,Kehagias:2016zry , Hamada:2017gdg} and the possible interplay between symmetries \cite{Anninos:2010zf,Hinterbichler:2013dpa,Horn:2014rta,Compere:2019bua,Ashtekar:2014zfa,Balakrishnan:2019zxm}, memory effects \cite{Bieri:2015jwa , Chu:2015yua, Tolish:2016ggo , Chu:2016qxp , Chu:2016ngc , Hamada:2017gdg,Bieri:2017vni, Chu:2019ssw} and consistency conditions for correlation functions \cite{Maldacena:2002vr,Creminelli:2004yq,Cheung:2007sv,Hinterbichler:2013dpa,Horn:2014rta,Hui:2018cag, Banerjee:2020dww}, playing a similar role than the BMS$_n$ group in asymptotically flat spacetimes (see \textit{e.g.} \cite{Strominger:2017zoo, Ashtekar:2014zsa,Ashtekar:2018lor,Compere:2018aar,Ruzziconi:2019pzd} for partial reviews).

The paper is organized as follows. In section \ref{sec:Gravity in Fefferman-Graham gauge}, we define the notion of Al(A)dS$_n$ spacetimes, derive the solution space and compute the residual gauge diffeomorphism algebra and its action on the solution space (see appendices \ref{Review of the solution space}, \ref{app:Algebra of residual gauge diffeomorphisms} and \ref{Variations of the solution space} for the detailed computations). The section \ref{sec:Holographic renormalization} is devoted to a review of the holographic renormalization procedure and how to bring the counter-terms at the level of the symplectic form in the covariant phase space formalism (see also appendix \ref{app:Holographic renormalization}). In section \ref{sec:Charge algebra in asymptotically locally}, we derive the charges associated with the most general residual gauge diffeomorphisms and show that they satisfy an algebra involving a field-dependent $2$-cocycle (see also appendix \ref{app:Surface charges}, \ref{Derivation of the charge algebra}, and \ref{New massive gravity}). Next, in section \ref{Application to more restrictive boundary conditions}, we apply our general framework to more restrictive set-ups that include Dirichlet, Neumann and some leaky boundary conditions. Finally, in section \ref{sec:Comments}, we comment on our results and elaborate on possible implications.

\paragraph{Conventions} We write $\mathscr{M}$ the spacetime manifold and $\mathscr{I}$ its boundary. The bulk spacetime dimension is $n = d+1$, where $d$ is the boundary dimension ($n \ge 3$ and $d \ge 2$). We write $\Lambda = - \eta \frac{d(d-1)}{2 \ell^2}$ the cosmological constant, where $\eta = - \text{sgn}(\Lambda)$ is minus the sign of the cosmological constant. The Riemann tensor is defined using the conventions of \cite{Misner_1973} such that $R<0$ for AdS$_{d+1}$. The Starobinsky/Fefferman-Graham spacetime coordinates are given by $x^\mu = (\rho, x^a)$, $a=1 \ldots d$, where $\rho\geq 0$ has units of inverse of length and the other coordinates $x^a$ are dimensionless. The boundary $\mathscr I$, located at $\rho = 0$, is timelike if $\eta = +1$ and spacelike if $\eta = -1$. We will denote by $\mathscr{I}_{\text{AdS}}$ the timelike boundary of AdS and by $\mathscr{I}^+_{\text{dS}}$ the future spacelike boundary of dS (see Figure \ref{fig:Conservative vs leaky}). The numerically invariant Levi-Civita symbol $\varepsilon_{\mu_1 \dots \mu_{d+1}} = \varepsilon_{[\mu_1 \dots \mu_{d+1}]}$, $\varepsilon_{1 \dots d+1} = 1$, is such that $\varepsilon_{\rho a_1 \dots a_d}  = \varepsilon_{a_1 \dots a_d}$. A codimension $k$ form in $d+1$ dimensions is written as $\bm \alpha = \alpha^{\mu_1 \dots \mu_k} (\D^{d+1-k} x)_{\mu_1 \dots \mu_k}$ where
\begin{equation}
(\D^{d+1-k} x)_{\mu_1 \dots \mu_k} = \frac{1}{k!(d+1-k)!}\varepsilon_{\mu_1 \dots \mu_k \nu_1 \ldots \nu_{d+1-k}} \D x^{\nu_1} \wedge \dots \wedge \D x^{d+1-k}.
\end{equation} 
Consistently, we also define $(\D^d x) \equiv (\D^{d}x)_{\rho}$ and $(\D^{d-1}x) \equiv 2\ell (\D^{d-1}x)_{\rho t}$ where $t\equiv \ell x^1$. When we perform integration on the manifold, we denote $\int_{\mathscr M} (...) = \int_0^\infty \D\rho' \int_{\rho=\rho'} (\D^d x) (...)$. This convention sets the lower bound of the bulk integral to be the boundary $\rho=0$. The integration measure $(\D^d x)$ should thus be understood as a measure on the hypersurface at fixed $\rho=\rho'$. In the Starobinsky/Fefferman-Graham expansion, we denote by $\Phi^{(k)}$ the coefficients in the $\rho$-expansion of some field $\Phi$ regardless to the dimension, and by $\Psi^{[d]}$ some field $\Psi$ specific to the dimension $d$. Finally we define the variation $\delta_\xi$ of the metric tensor such that $\delta_\xi g_{\mu\nu} = +\mathcal L_\xi g_{\mu\nu}$, which is the same sign convention used in \cite{Compere:2020lrt} but the opposite of the one used in \cite{Barnich:2011mi}. Accordingly, the Barnich-Troessaert bracket is modified by a global sign with respect to \cite{Barnich:2011mi}.

\section{Gravity in Starobinsky/Fefferman-Graham gauge}
\label{sec:Gravity in Fefferman-Graham gauge}

In this section, we review some standard results concerning gravity in the Starobinsky/Fefferman-Graham gauge and derive the algebra of the residual gauge diffeomorphisms and its action on the solution space. This allows us to install our conventions and define the general framework in which we will construct the phase space.   

\subsection{Gauge fixing and solution space}

In this work, we consider the most generic metric written in the Starobinsky/Fefferman-Graham gauge \cite{Starobinsky:1982mr,Fefferman:1985aa}
\begin{equation}
\D s^2 =\eta \frac{\ell^2}{\rho^2}\D\rho^2 + \gamma_{ab}(\rho,x^c) \D x^a \D x^b. \label{FG gauge}
\end{equation}
We only assume the fall-off condition $\gamma_{ab} = \mathcal{O}(\rho^{-2})$ that allows for conformal compactification. The analysis that follows does not impose any further boundary condition (except in section \ref{Application to more restrictive boundary conditions} where we will restrict our considerations to particular sets of boundary conditions to relate our results to previous analyses). We say that a spacetime is asymptotically locally (A)dS$_{d+1}$ (or Al(A)dS$_{d+1}$ for short) if the associated metric written in the Starobinsky/Fefferman-Graham gauge \eqref{FG gauge} obeys the fall-off condition and satisfies the Einstein equations with $\Lambda \neq 0$. The general asymptotic expansion that solves the Einstein equations reads as \cite{deHaro:2000vlm}
\begin{equation}
\gamma_{ab} = \rho^{-2} g_{ab}^{(0)} + g_{ab}^{(2)}+ \dots + \rho^{d-2} g_{ab}^{(d)} + \rho^{d-2} \ln \rho^2 \tilde{g}_{ab}^{[d]}    +  \mathcal{O}(\rho^{d-1}) 
\label{preliminary FG}
\end{equation} where the logarithmic term appears only for even $d$. This expansion is completely determined by specifying $g_{ab}^{(0)}$ and ${g}_{ab}^{(d)}$ (see appendix \ref{Review of the solution space}). We call $g^{(0)}_{ab}$ the boundary metric. Strictly speaking, it corresponds to a representative of the conformal class induced on the spacetime boundary through the conformal compactification procedure (see \textit{e.g.} \cite{Papadimitriou:2005ii}). For future purposes, we define the holographic stress-energy tensor as
\begin{equation}
\boxed{
T_{ab}^{[d]} = \frac{d}{16 \pi G} \frac{\eta}{\ell} \left( g_{ab}^{(d)} + X^{[d]}_{ab}[g^{(0)}] \right)
\label{holographic stress-energy tensor}
}
\end{equation}
where $X_{ab}^{[d]} [g^{(0)}]$ is a function which vanishes for $d$ odd \cite{deHaro:2000vlm, Balasubramanian:1999re}. For $d=2$, it is explicitly given by 
\begin{equation}
X^{[2]}_{ab}[g^{(0)}] = -g_{ab}^{(0)} (g^{cd}_{(0)} g^{(2)}_{cd}) , \label{X2}
\end{equation} while for $d=4$, we have
\begin{equation}
X^{[4]}_{ab}[g^{(0)}] =  - \frac{1}{8} g_{ab}^{(0)} [(g^{cd}_{(0)} g_{cd}^{(2)})^2 - g_{(2)}^{cd} g^{(2)}_{cd}] - \frac{1}{2} g^{(2)}_{ac} g^{c}_{(2)b} + \frac{1}{4} g_{ab}^{(2)} (g^{cd}_{(0)} g^{(2)}_{cd} ) . \label{X4}
\end{equation} Here and in the following, indices are lowered and raised by the induced spacetime boundary metric $g^{(0)}_{ab}$ and its inverse $g_{(0)}^{ab}$, \textit{e.g.} $g_{(2)}^{cd}= g^{ca}_{(0)} g^{db}_{(0)}  g^{(2)}_{ab}$. The Einstein equations imply
\begin{equation}
D^a T_{ab}^{[d]} = 0 
\label{divergence free}
\end{equation} where $D_a$ is the Levi-Civita connection with respect to $g^{(0)}_{ab}$. As discussed in appendix \ref{Review of the solution space}, the term $X^{[d]}_{ab}[g^{(0)}]$ in the definition \eqref{holographic stress-energy tensor} is precisely such that the holographic stress-energy tensor is divergence-free. For odd $d$, the Einstein equations also yield
\begin{equation}
\mathcal{T}^{[2k+1]} \equiv  g_{(0)}^{ab} T_{ab}^{[2k+1]} = 0, \, \forall \, k \in \mathbb N_0. \label{trace even}
\end{equation} Furthermore, for $d=2$, we have 
\begin{equation}
\mathcal{T}^{[2]} \equiv g_{(0)}^{ab} T_{ab}^{[2]} = \frac{c}{24 \pi} R^{(0)} = \frac{\ell}{\sqrt{|g^{(0)}|}} L_{EH}[g^{(0)}]  \label{trace 3d}
\end{equation} where $c = \frac{3 \ell}{2G}$ is the Brown-Henneaux central charge inducing the conformal anomaly in two dimensions \cite{Brown:1986nw} and $L_{EH}$ is the Einstein-Hilbert Lagrangian density. For $d=4$, we have
\begin{equation}
\begin{split}
\mathcal{T}^{[4]} \equiv g_{(0)}^{ab} T_{ab}^{[4]} =& - \frac{1}{16 \pi G}\frac{\eta}{\ell}\left[(g^{cd}_{(0)} g_{cd}^{(2)})^2 - g_{(2)}^{cd} g^{(2)}_{cd}\right] = \frac{\eta\,\ell^3}{64 \pi G} \left( R_{ab}^{(0)} R^{ab}_{(0)} - \frac{R_{(0)}^2}{3} \right) \\
=& \frac{\eta\,\ell^3}{4\sqrt{|g^{(0)}|}} \left( L_{QCG(1)}[g^{(0)}] - \frac{1}{3} L_{QCG(2)}[g^{(0)}] \right) 
\end{split} \label{trace 4d}
\end{equation}  where we used that $g^{ab}_{(0)} g^{(4)}_{ab} = \frac{1}{4} g_{(2)}^{ab} g^{(2)}_{ab}$ on-shell. Again, this reproduces the conformal anomaly in four dimensions \cite{Henningson:1998gx}. Interestingly, this expression reveals the Lagrangian densities $L_{QCG(1)}[g^{(0)}]$ and $L_{QCG(2)}[g^{(0)}]$ of the quadratic curvature gravity on the boundary \cite{Salvio:2018crh , Stelle:1977ry}. The expressions of $L_{EH}$, $L_{QCG(1)}$ and $L_{QCG(2)}$ are included in appendix \ref{New massive gravity}.

\subsection{Residual gauge diffeomorphisms}

The infinitesimal diffeomorphisms preserving the Starobinsky/Fefferman-Graham gauge are generated by vector fields $\xi = \xi^\rho \partial_\rho + \xi^a \partial_a$ satisfying
$\mathcal{L}_\xi g_{\rho \rho} = 0$, $\mathcal{L}_\xi g_{\rho a} = 0$. The first condition leads to the equation 
\begin{equation}
\partial_\rho \xi^\rho = \frac{1}{\rho} \xi^\rho, \label{eq:CstVector1}
\end{equation}
which can be solved for $\xi^\rho$ as 
\begin{equation}
\xi^\rho = \sigma (x^a) \rho \label{AKV 1}
\end{equation}
where $\sigma (x^a)$ is a possibly field-dependent arbitrary function of the boundary coordinates that parametrizes the Weyl rescalings. The second condition leads to the equation 
\begin{equation}
\rho^2 \gamma_{ab} \partial_\rho \xi^b + \eta\ell^2\partial_a \xi^\rho =0, \label{eq:CstVector2}
\end{equation}
which can be solved for $\xi^a$ as 
\begin{equation}
\xi^a = \bar \xi ^a (x^b) -\eta\ell^2\partial_b \sigma \int_0^\rho \frac{\D\rho'}{\rho'} \gamma^{ab}(\rho', x^c)
\label{AKV 2}
\end{equation} where $\bar \xi ^a (x^b)$ is a possibly field-dependent arbitrary vector field on the spacetime boundary. The subset of residual gauge diffeomorphisms parametrized by $\sigma$ are referred as the Penrose-Brown-Henneaux (PBH) transformations \cite{Penrose:1986ca, Brown:1986nw , Imbimbo:1999bj}.

Using the modified Lie bracket that takes into account the field-dependence of the vectors fields \cite{Schwimmer:2008yh, Barnich:2010eb},
\begin{equation}
[\xi_1, \xi_2 ]_\star = [\xi_1, \xi_2 ] - \delta_{\xi_1} \xi_2 + \delta_{\xi_2} \xi_1 , 
\label{modified bracket}
\end{equation} we show in appendix \ref{app:Algebra of residual gauge diffeomorphisms} that the residual gauge diffeomorphisms satisfy 
\begin{equation}
\boxed{
\begin{gathered}
{}[\xi(\sigma_1,\bar \xi_{1}^a),\xi(\sigma_2,\bar \xi_{2}^a)]_\star =  \xi (\hat{\sigma},\hat{\bar \xi}^a) ,\\
 \text{with} \ 
	\left\lbrace
	\begin{gathered}
		\,\,\hat \sigma = \bar{\xi}_{1}^a\partial_a \sigma_2  - \delta_{\xi_1} \sigma_2 - (1 \leftrightarrow 2 ), \\
        \,\,\hat{\bar \xi}^a =  \bar \xi^b_{1} \partial_b \bar \xi^a_{2} - \delta_{\xi_1} \bar \xi_{2}^a - (1 \leftrightarrow 2).
	\end{gathered}
	\right.
\end{gathered}
}
\label{eq:VectorAlgebra}
\end{equation} 
This generalizes the result obtained for $d=3$ in \cite{Compere:2020lrt} to arbitrary $d \ge 2$. This also extends the analysis of \cite{Schwimmer:2008yh} where the modified Lie bracket \eqref{modified bracket} was applied to the subclass of PBH diffeomorphisms. In this derivation, we did not assume that the residual gauge diffeomorphism parameters $\sigma$ and $\bar \xi^a$ are field-independent, which implies the presence of the terms $\delta_{\xi} \sigma$ and $\delta_{\xi} \bar \xi^a$ in \eqref{eq:VectorAlgebra}. Indeed, as explained in section \ref{Application to more restrictive boundary conditions}, the restriction of our general analysis to particular sets of boundary conditions may require field-dependence of these parameters. Finally, if one assumes that the parameters are field-independent at this stage ($\delta_{\xi} \sigma =0$ and $\delta_{\xi} \bar \xi^a =0$), then the commutation relations \eqref{eq:VectorAlgebra} reduce to those of the semi-direct sum $\text{Diff}(\mathscr{I}) \loplus \mathbb{R}$ where Diff($\mathscr{I}$) denotes the diffeomorphisms on the boundary $\mathscr{I}$, parametrized by $\bar{\xi}^a$, and $\mathbb{R}$ denotes the abelian Weyl rescalings on the boundary, parametrized by $\sigma$.

\subsection{Variations of the solution space}

Under the infinitesimal residual gauge diffeomorphisms \eqref{AKV 1} and \eqref{AKV 2}, the boundary metric transforms as
\begin{equation}
\boxed{
\delta_\xi g^{(0)}_{ab} = \mathcal{L}_{\bar\xi} g^{(0)}_{ab} - 2 \sigma g_{ab}^{(0)} . } \label{eq:action solution space}
\end{equation} This justifies the names of boundary diffeomorphisms and Weyl rescalings for the symmetry parameters $\bar{\xi}^a$ and $\sigma$, respectively. The holographic stress-energy tensor transforms as
\begin{equation}
\boxed{
\delta_{\xi} T_{ab}^{[d]} = \mathcal{L}_{\bar\xi}  T_{ab}^{[d]} + (d-2) \sigma  T_{ab}^{[d]}  + A_{ab}^{[d]}[\sigma] }
\label{eq:action solution space 2}
\end{equation} where $A_{ab}^{[d]}[\sigma]$ denotes the inhomogeneous part of the transformation due to Weyl rescalings. We have $A_{ab}^{[2k+1]}[\sigma] = 0$ for $k \in\mathbb N_0$. For $d = 2$, we find
\begin{equation}
A_{ab}^{[2]}[\sigma] = - \frac{\ell}{8 \pi G} (D_a D_b\sigma - g^{(0)}_{ab} D^c D_c \sigma ), 
\end{equation} while for $d=4$, we obtain
\begin{equation}
\begin{split}
A_{ab}^{[4]}[\sigma] = \frac{\eta}{4\pi G\ell} &\left[ 2 \sigma \tilde{g}^{[4]}_{ab} +  \frac{\ell^4}{4} D^c \sigma \Big( D_c R_{ab}^{(0)} - D_{(a} R_{b)c}^{(0)}- \frac{1}{6} D_c R^{(0)} g_{ab}^{(0)} \Big) \right. \\ 
&\phantom{\Big[} + \frac{\ell^4}{24} D_{(a} \sigma D_{b)} R^{(0)} + \frac{\ell^4}{12} R^{(0)} \Big( D_a D_b \sigma - g_{ab}^{(0)} D^c D_c \sigma  \Big) \\
&\phantom{\Big[} \left. + \frac{\ell^4}{8} \Big(  R_{ab}^{(0)} D^c D_c \sigma - 2 R_{c(a}^{(0)} D^c D_{b)} \sigma + g^{(0)}_{ab} R_{cd}^{(0)} D^c D^d \sigma  \Big) \right]  
\end{split}
\end{equation} where $\tilde{g}^{[4]}_{ab}$ is given explicitly in equation \eqref{tilde 4}. In appendix \ref{Variations of the solution space}, we provide some intermediate steps to obtain these variations. 

From general considerations \cite{Henneaux:1992ig , Barnich:2010xq , Barnich:2018gdh , Barnich:2007bf, Compere:2018aar , Ruzziconi:2019pzd}, the action on the solution space prescribed by \eqref{eq:action solution space} and \eqref{eq:action solution space 2} forms a representation of the symmetry algebra \eqref{eq:VectorAlgebra}, namely
\begin{equation}
[\delta_{\xi_1} ,  \delta_{\xi_2}] (g^{(0)}_{ab}, T^{[d]}_{ab})  = - \delta_{[\xi_1, \xi_2]_\star} (g^{(0)}_{ab}, T^{[d]}_{ab})  
\end{equation} where $[\delta_{\xi_1} ,  \delta_{\xi_2}] = \delta_{\xi_1}  \delta_{\xi_2} - \delta_{\xi_2}  \delta_{\xi_1}$.

\section{Holographic renormalization}
\label{sec:Holographic renormalization}

In this section, we review the holographic renormalization procedure that yields a finite on-shell action in Al(A)dS$_{d+1}$ spacetimes \cite{deHaro:2000vlm}. More details of this procedure in our framework and conventions can also be found in appendix \ref{app:Holographic renormalization}. We then report the counter-terms at the level of the symplectic structure to obtain finite expression on the phase space \cite{Papadimitriou:2005ii , Compere:2008us}.  

\subsection{Renormalized action}

The boundary counter-terms to be added to the Einstein-Hilbert action in Al(A)dS$_{d+1}$ spacetimes were obtained in \cite{deHaro:2000vlm} by requiring that the total action is finite on-shell. It was subsequently shown in \cite{Papadimitriou:2005ii} that these counter-terms were precisely those needed to write a variational principle consistent with the formulation of the boundary data in terms of boundary conformal classes. In the present paper, we stick to the former approach which is sufficient to discuss the renormalization at the level of the presymplectic structure.

The renormalized action for General Relativity without matter in Al(A)dS$_{d+1}$ spacetimes \cite{deHaro:2000vlm , Bianchi:2001kw} is generically given by 
\begin{equation}
S_{ren} = \int_{\mathscr{M}}  \bm L_{EH} + \int_{\mathscr{I}} \bm  L_{GHY}+  \int_{\mathscr{I}}   \bm L_{ct} + \int_{\mathscr{I}} \bm L_\circ .
\label{renormalized action}
\end{equation} The first term in the right-hand side of \eqref{renormalized action} is the Einstein-Hilbert action whose Lagrangian $(d+1)$-form is
\begin{equation}
\bm L_{EH}[g] = \frac{1}{16 \pi G} \sqrt{-g} \left( R[g] + \eta \frac{d(d-1)}{\ell^2} \right) \D^{d+1}x . \label{LEH}
\end{equation} The second term is the Gibbons-Hawking-York boundary term
\begin{equation}
\bm L_{GHY}[\gamma] = \frac{1}{8\pi G} \eta \sqrt{|\gamma|} K \D^d x  \label{LGHY}
\end{equation} that allows to have an action stationary on solutions (\textit{i.e.} $\delta S_{ren} = 0$ on-shell) in the particular case of Dirichlet boundary conditions ($\delta (\rho^2 \gamma_{ab})|_{\mathscr{I}} = \delta g^{(0)}_{ab} = 0$). It involves the second fundamental form $K_{ab} = \frac{1}{2} \mathcal{L}_N \gamma_{ab} = \nabla_{(a} N_{b)}$ of $\mathscr I$ and its trace $K = \gamma^{ab} K_{ab}$, the extrinsic curvature. These objects are built up with the outward normal vector $N = N^\mu \partial_\mu = - \sqrt{|g^{\rho\rho}|} \partial_\rho$, $N^\mu N_\mu = \eta$, which is the natural background structure induced by the Starobinsky/Fefferman-Graham foliation \eqref{FG gauge}. The two last pieces of \eqref{renormalized action} are meant respectively to renormalize the on-shell action, and adjust the finite piece of the variational principle as we discuss now. Following the procedure reviewed and detailed in appendix \ref{app:Holographic renormalization}, one can control the divergences of the on-shell action by introducing an infrared cut-off $\epsilon>0$ (called the regulator), and perform the integration towards the boundary up to $\rho=\epsilon$. This defines the regularized action
\begin{equation}
S_{reg}^\epsilon = \int_{\rho \ge \epsilon} \bm L_{EH} + \int_{\rho = \epsilon}  \bm L_{GHY} ,
\end{equation}  
whose on-shell evaluation displays the following radial divergences:
\begin{equation}
S_{reg}^\epsilon = \frac{\eta}{16\pi G\ell} \int_{\rho = \epsilon} \D^d x  \sqrt{|g^{(0)}|} \left(\epsilon^{-d} a_{(0)} +  \epsilon^{-d+2} a_{(2)} + \cdots + \epsilon^{-2} a_{(d-2)} - \ln \epsilon^2 ~ \tilde{a}_{[d]}\right) + \mathcal{O}(\epsilon^0).
\end{equation} 
The coefficients $a_{(i)}$, $\forall\,i\leq (d-2)$, and $\tilde a_{[d]}$ are local covariant expressions involving the boundary metric $g^{(0)}_{ab}$ and its curvature tensor. Up to the order of interest for us, we have
\begin{equation}
a_{(0)} = 2 (d-1), \quad a_{(2)} = \frac{(d-4)(d-1)}{d-2} g_{(0)}^{ab} g^{(2)}_{ab} \quad (d > 2)
\end{equation} and 
\begin{equation}
\tilde{a}_{[2]} = - g^{ab}_{(0)} g^{(2)}_{ab} , \quad \tilde{a}_{[4]} = -\frac{1}{2}\left[ (g_{(0)}^{ab} g^{(2)}_{ab})^2 - g_{(2)}^{ab} g^{(2)}_{ab} \right].
\end{equation}
The coefficient $\tilde{a}_{[d]}$ appears only for even $d$ and is proportional to the conformal anomaly, since
\begin{equation}
\mathcal T^{[2]} = \frac{\eta}{8\pi G\ell}\tilde{a}_{[2]} , \quad \mathcal T^{[4]} = \frac{\eta}{8\pi G\ell}\tilde{a}_{[4]} .
\end{equation}
The counter-term to be added to the action to make it finite on-shell is given by
\begin{equation}
S_{ct}^\epsilon =  \int_{\rho = \epsilon} \bm L_{ct}, \quad \bm L_{ct} =  \frac{1}{16\pi G}\frac{\eta}{\ell} \sqrt{|\gamma|} \Big( -2(d-1) - \frac{\eta \ell^2}{(d-2)}R[\gamma] + \ldots + \ln \epsilon^2 ~\tilde a_{[d]}   \Big) \D^{d}x 
\end{equation} where the second term arises only for $d \ge 3$ and the dots indicate the presence of higher curvature terms for $d \ge 5$.

Finally, depending on the motivations and the particular boundary conditions under consideration, one can always add an additional term to the action 
\begin{equation}
S^\epsilon_\circ = \int_{\rho = \epsilon} \bm L_{\circ}[\gamma]
\end{equation} that is finite on-shell, \textit{i.e.} $\bm L_{\circ} = \mathcal{O}(\epsilon^0)$, and covariant with respect to the metric $\gamma_{ab}$. As detailed in appendix \ref{app:Holographic renormalization}, on the boundary, we choose this Lagrangian to be 
\begin{equation}
\bm L_\circ[g^{(0)}]  = \kappa_{[d]} \sqrt{|g^{(0)}|} \tilde{a}_{[d]} \D^d x, \quad \kappa_{[2k+1]} = 0, \, \kappa_{[2]} = 0, \, \kappa_{[4]} = \frac{3}{2} ,
\end{equation} which allows to make the definitions of the holographic stress-energy tensor \eqref{holographic stress-energy tensor} and \eqref{Tab general} coincide. Indeed, the renormalized action \eqref{renormalized action} can be written as
\begin{equation}
S_{ren} = \lim_{\epsilon\to 0}(S_{reg}^\epsilon + S_{ct}^\epsilon + S^\epsilon_\circ)    \label{ren action principle}
\end{equation} and the holographic stress-energy tensor defined in \eqref{holographic stress-energy tensor} can be derived from it through \cite{deHaro:2000vlm} 
\begin{equation}
T_{ab}^{[d]} = -\frac{2}{\sqrt{|g^{(0)}|}} \frac{\partial S_{ren}}{\partial g_{(0)}^{ab}}   \label{Tab general}
\end{equation}
where $S_{ren}$ is evaluated on-shell before taking the functional derivative with respect to $g^{(0)}_{ab}$.

\subsection{Renormalized symplectic structure and variational principle}
\label{sec:Renormalized symplectic structure and variational principle}

Following the prescription described in references \cite{Papadimitriou:2005ii , Compere:2008us}, we use the holographically renormalized action \eqref{renormalized action} to remove the divergences of the presymplectic potential and fix the ambiguities of the covariant phase space formalism \cite{Lee:1990nz , Wald:1993nt , Iyer:1994ys , Wald:1999wa} (see \textit{e.g.} \cite{Compere:2018aar , Ruzziconi:2019pzd} for reviews). We write $\bm \Theta_{EH}$, $\bm \Theta_{GHY}$, $\bm \Theta_{ct}$ and $\bm \Theta_\circ$ the presymplectic potentials of $\bm L_{EH}$, $\bm L_{GHY}$, $\bm L_{ct}$ and $\bm L_{\circ}$, respectively. More explicitly, we have
\begin{equation}
\begin{split}
&\delta \bm L_{EH} = \frac{\delta \bm L_{EH}}{\delta g^{\mu\nu}}\delta g^{\mu\nu} + \D\bm\Theta_{EH}   , \quad \delta \bm L_{GHY} = \frac{\delta \bm L_{GHY}}{\delta \gamma^{ab}}\delta \gamma^{ab} + \D\bm\Theta_{GHY}, \\
&\delta \bm L_{ct} = \frac{\delta \bm L_{ct}}{\delta \gamma^{ab}}\delta \gamma^{ab} + \D\bm\Theta_{ct}    , \quad \delta \bm L_{\circ} = \frac{\delta \bm L_{\circ}}{\delta \gamma^{ab}}\delta \gamma^{ab} + \D\bm\Theta_{\circ}   
\end{split} \label{defs of presymplectic}
\end{equation} where it can be shown that $\bm\Theta_{GHY} =0$ canonically. The renormalized presymplectic potential is obtained by taking a variation of the renormalized Lagrangian $\bm L_{ren}$ associated with \eqref{renormalized action}. Using Stokes theorem to rewrite $\bm L_{ren}$ as a bulk Lagrangian and taking into account the orientation of the normal, we obtain explicitly
\begin{equation}
\begin{split}
\delta\bm L_{ren} &= \delta \bm L_{EH}-\D( \delta \bm L_{GHY} + \delta \bm L_{ct} + \delta \bm L_{\circ})\\
&= \frac{\delta \bm L_{EH}}{\delta g^{\mu\nu}}\delta g^{\mu\nu} + \D\bm\Theta_{EH} - \D \left( \frac{\delta (\bm L_{GHY} + \bm L_{ct} + \bm L_{\circ})}{\delta \gamma^{ab}}\delta \gamma^{ab}  +\D (\bm \Theta_{GHY} + \bm \Theta_{ct}+ \bm \Theta_{\circ} ) \right) \\
&= \frac{\delta \bm L_{EH}}{\delta g^{\mu\nu}}\delta g^{\mu\nu} + \D \left( \bm\Theta_{EH} - \frac{\delta (\bm L_{GHY} + \bm L_{ct} + \bm L_{\circ})}{\delta \gamma^{ab}}\delta \gamma^{ab} \right)
\end{split}
\end{equation}
where we used \eqref{defs of presymplectic} in the second equality and $\D^2 = 0$ in the third equality. Taking into account
\begin{equation}
\delta \bm L_{ren} = \frac{\delta \bm L_{ren}}{\delta g^{\mu\nu}}\delta g^{\mu\nu} + \D\bm\Theta_{ren} , \qquad \frac{\delta \bm L_{ren}}{\delta g^{\mu\nu}} = \frac{\delta \bm L_{EH}}{\delta g^{\mu\nu}} ,
\end{equation} we can identify the (canonical) renormalized presymplectic potential as
\begin{equation}
\bm\Theta_{ren} = \bm\Theta_{EH} - \frac{\delta (\bm L_{GHY} + \bm L_{ct} + \bm L_{\circ})}{\delta \gamma^{ab}}\delta \gamma^{ab}
\end{equation} or, using \eqref{defs of presymplectic}, 
\begin{equation}
\boxed{
\bm \Theta_{ren} [g ; \delta g] \equiv \bm \Theta_{EH} - \delta \bm L_{GHY} - \delta \bm L_{ct} - \delta \bm L_\circ + \D \bm \Theta_{ct} + \D \bm \Theta_{\circ} . }
\label{renormalized presympelctic}
\end{equation}
In the framework of the covariant phase space methods, this renormalization procedure involves the two types of ambiguities arising in the formalism. Indeed, adding a boundary term to the Einstein-Hilbert Lagrangian $\bm A \equiv \bm{L}_{GHY} + \bm{L}_{ct} + \bm{L}_\circ$ modifies the presymplectic potential as $\bm{\Theta}_{EH} \to \bm{\Theta}_{EH} - \delta \bm{A}$. Furthermore, the presymplectic potential is defined up to an exact $d$-form, $\bm{\Theta}_{EH} \to \bm{\Theta}_{EH} + \D \bm{Y}$, which is taken as $\bm Y \equiv \bm \Theta_{ct} + \bm \Theta_{\circ} $, in such a way that the remaining finite piece is linear in $\delta g^{(0)}_{ab}$ and only involves the holographic stress-tensor, thanks to \eqref{Tab general}. These two ambiguities reproduce precisely \eqref{renormalized presympelctic}. The resulting pull-back on $\mathscr I$ of the renormalized presymplectic potential reads as
\begin{equation}
\boxed{
\bm \Theta_{ren}[g;\delta g]\Big|_{\mathscr I} = - \frac{1}{2} \sqrt{|g^{(0)}|}  T^{ab}_{[d]} \delta g^{(0)}_{ab} (\D^{d}x)  .
} \label{Theta ren pullback}
\end{equation}
Notice that \eqref{Theta ren pullback} encodes the variation of the action when evaluated on a solution, namely
\begin{equation}
\delta S_{ren} = -\int_{\mathscr{I}} \bm \Theta_{ren} [g ; \delta g]\Big|_{\mathscr{I}}  
\label{var principle theta}
\end{equation} where we have considered only the spacetime boundary $\mathscr{I}$. The renormalized presymplectic current is defined as
\begin{equation}
\bm \omega_{ren} [g ; \delta_1 g, \delta_2 g] = \delta_1 \bm \Theta_{ren} [g ; \delta_2 g] - \delta_2 \bm \Theta_{ren} [g ; \delta_1 g]  .
\end{equation}
When pulled back on $\mathscr I$, we obtain 
\begin{equation}
\boxed{
\bm \omega_{ren} [g ; \delta_1 g, \delta_2 g] \Big|_{\mathscr I} = - \frac{1}{2} \delta_1 \left( \sqrt{|g^{(0)}|}  T^{ab}_{[d]} \right) \delta_2 g^{(0)}_{ab} \, (\D^{d}x) - (1 \leftrightarrow 2) . 
}
\label{renormalized presymplectic current}
\end{equation} As discussed in sections \ref{sec:Charge algebra in asymptotically locally} and \ref{Application to more restrictive boundary conditions}, this formula encodes the flux of charges through the spacetime boundary $\mathscr{I}$. While this is natural to keep this flux as it is in AldS spacetimes, conservative boundary conditions in AlAdS would impose further restrictions to set $\bm \omega_{ren} [g ;\delta_1 g , \delta_2 g]|_{\mathscr{I}}= 0$, in order to obtain conserved charges. In that case, one can always find an action that is stationary on solutions by using the freedom to add finite boundary terms \cite{Papadimitriou:2005ii}. In particular, for Dirichlet boundary conditions that are discussed in section \ref{Dirichlet}, no additional term is required. 

Here, we pursue the analysis without imposing that the flux vanishes at the boundary. This yields an open gravitational system with some leaks at the spacetime boundary and the action cannot be made stationnary on solutions. In AlAdS spacetimes, starting from the theory defined by the action \eqref{renormalized action} and with associated presymplectic potential \eqref{Theta ren pullback}, it is natural to interpret the system as gravity with $\Lambda <0$ coupled with some external sources encoded in the fluctuations of the boundary metric \cite{Compere:2008us , Troessaert:2015nia}. Turning off the sources (\textit{i.e.} setting $\delta g^{(0)}_{ab} =0$) renders a closed system. This point of view will be further described in section \ref{sec:Comments}. As we will see, the analysis of asymptotics with such leaky boundary conditions does not require to specify the precise nature of the environment. The latter could for example be taken as an asymptotically flat region beyond the conformal boundary, as depicted in Figure \hyperlink{fig:AdS}{1.(a)} (see \cite{Almheiri:2019yqk , Almheiri:2019qdq , Almheiri:2020cfm}). Specifying the nature of the reservoir would deliver an interpretation for the fluctuations of the boundary structure that play the role of external sources in our analysis. In particular, if the system and the environment form together a closed system, the stationarity of the on-shell action can be restored.

Notice that the picture described here for Al(A)dS spacetimes with leaky boundary conditions is very similar to the one of asymptotically flat spacetimes at null infinity. Indeed, in the latter case, the symplectic flux is non-vanishing at infinity due to the leaks of gravitational waves throughout the null spacetime boundary \cite{Sachs:1962wk , 1962RSPSA.269...21B , Penrose:1965am , Newman:1968uj , Ashtekar:2014zsa}. This leads to an open system for which the action is not stationary on-shell (see \textit{e.g.} \cite{Compere:2018ylh} for a recent discussion). The analogy between Al(A)dS and asymptotically flat spacetimes will be further explored in section \ref{sec:Comments}. 

\section{Charge algebra in asymptotically locally \texorpdfstring{(A)dS$_{d+1}$}{(A)dS d+1} spacetimes}
\label{sec:Charge algebra in asymptotically locally}

This section contains the main results of the paper. We derive the infinitesimal charges of Al(A)dS$_{d+1}$ spacetimes. The charges associated with boundary diffeomorphisms are generically non-vanishing, while those associated with Weyl rescalings are non-vanishing only in odd spacetime dimension. We interpret this fact from the point of view of the dual theory as related to the presence of Weyl anomalies. Finally, we derive the charge algebra using the Barnich-Troessaert bracket \cite{Barnich:2011mi} and show that a field-dependent $2$-cocycle shows up in odd spacetime dimensions.

\subsection{Infinitesimal charges}

In the covariant phase space formalism \cite{Lee:1990nz , Wald:1993nt , Iyer:1994ys , Wald:1999wa , Barnich:2001jy, Barnich:2003xg, Barnich:2007bf} (see also \cite{Compere:2018aar , Ruzziconi:2019pzd} for reviews), the infinitesimal charges are obtained by integrating the codimension $2$ forms $\bm k_{\xi ,ren}[g; \delta g ]$ associated with residual gauge diffeomorphisms $\xi$, on a codimension $2$ section $S_\infty \equiv \lbrace \rho,t=\text{constant}\rbrace$ of $\mathscr{I}$ as follows:
\begin{equation}
\ndelta H_\xi [g] = \int_{S_\infty} \bm k_{\xi,ren}[g; \delta g] =  \int_{S_\infty} (\D^{d-1} x) k^{\rho t}_{\xi,ren}[g; \delta g] 
\label{integration of krhoa}
\end{equation} Here $t \equiv \ell x^1$ denotes the first coordinate among the transverse Starobinsky/Fefferman-Graham coordinates (given in units of length), timelike if $\eta = 1$, spacelike if $\eta =-1$, and identified as the time evolution along the cylinder in the AdS case, and the sphere radius in the dS case. 

The expression \eqref{integration of krhoa} is a $1$-form on the solution space, which justifies the appellation ``infinitesimal charge''. The codimension $2$ form $\bm k_{\xi,ren}[g; \delta g ]$ is defined from the renormalized presymplectic current \eqref{renormalized presymplectic current} as
\begin{equation}
\boxed{ \D \bm k_{\xi,ren}[g; \delta g] = \bm \omega_{ren} [ g;\delta_\xi g, \delta g] } \quad \Rightarrow \quad \partial_a k_{\xi,ren}^{\rho a}[g; \delta g] = \omega_{ren}^\rho [ g;\delta_\xi g, \delta g]. \label{fundamental formula}
\end{equation} This defines $k_{\xi,ren}^{\rho a}[g; \delta g]$ up to total derivative terms, $k_{\xi,ren}^{\rho a}[g; \delta g] \to k_{\xi,ren}^{\rho a}[g; \delta g] +\partial_b M_\xi^{[\rho ab]}[g;\delta g]$, where $M_\xi^{[\rho ab]}[g; \delta g]$ are the components of a codimension $3$ form. This ambiguity does not play any role when the integration \eqref{integration of krhoa} on $S_\infty$ is performed. As shown in appendix \ref{app:Surface charges}, the right-hand side of \eqref{fundamental formula} is given explicitly by
\begin{equation}
\begin{split}
\omega^\rho_{ren} [g; \delta_\xi g, \delta g] &=  \delta\left(\sqrt{|g^{(0)}|} T^{ab}_{[d]}\right)D_a \bar\xi_b  - \frac{1}{2} \sqrt{|g^{(0)}|} \left( D_c \bar\xi^c\, T^{ab}_{[d]} + \mathcal L_{\bar\xi} T^{ab}_{[d]}\right)\delta g_{ab}^{(0)} \\
&\phantom{=}\, - \delta \left( \sqrt{|g^{(0)}|} \mathcal{T}^{[d]}  \right) \sigma  - \frac{1}{2} \sqrt{|g^{(0)}|} A^{ab}_{[d]}[\sigma] \delta g_{ab}^{(0)} + \mathcal{O}(\rho).
\end{split}
\end{equation} After a lengthy computation displayed in appendix \ref{app:Surface charges}, one obtains the explicit expression of the infinitesimal charges in Al(A)dS$_{d+1}$ spacetimes
\begin{equation}
\boxed{
\ndelta H_\xi [g] = \int_{S_\infty} (\D^{d-1} x)\left[ \delta \left( \sqrt{|g^{(0)}|} {g^{tc}_{(0)} T^{[d]}_{bc}} \right) \bar\xi^b - \frac{1}{2} \sqrt{|g^{(0)}|} \bar\xi^t T^{bc}_{[d]} \delta g_{bc}^{(0)} + W^{[d]t}_\sigma [g; \delta g] \right]  
}
\label{surface charge expression}
\end{equation} where $W^{[2k+1]t}_\sigma [g; \delta g] = 0$ ($k\in\mathbb N_0$). For $d= 2$, we have
\begin{equation}
W_\sigma^{[2]t} [g; \delta g] = -\frac{\ell}{16 \pi G} D_b \sigma \left[ \sqrt{|g^{(0)}|} \delta g^{t b}_{(0)} + 2 \delta \sqrt{|g^{(0)}|} g^{tb}_{(0)} \right] - \ell \sigma  \Theta^t_{EH} [g^{(0)}; \delta g^{(0)}] .
\end{equation} For $d= 4$, we find 
\begin{align}
W^{[4] t}_\sigma[g;\delta g] =& \frac{\eta \, \ell^3}{16 \pi G} \left[ \frac{1}{6} \sqrt{|g^{(0)}|} R^{(0)} D_b \sigma \delta g^{tb}_{(0)} + \frac{1}{3} R^{(0)} D^t \sigma \delta \sqrt{|g^{(0)}|}   \right. \nonumber \\
& \left. \quad\quad\quad - \frac{1}{2} R^{tc}_{(0)} D_c \sigma \delta \sqrt{|g^{(0)}|} + \frac{1}{4} \sqrt{|g^{(0)}|} R_{cb}^{(0)} D^t \sigma \delta g^{bc}_{(0)} - \frac{1}{2} \sqrt{|g^{(0)}|} R^{(0)t}_c D_b \sigma \delta g^{bc}_{(0)} \right] \nonumber \\
&- \eta \frac{\ell^3}{4} \sigma \left[  \Theta^t_{QCG(1)}[g^{(0)}; \delta g^{(0)}]  - \frac{1}{3} \Theta^t_{QCG(2)}[g^{(0)}; \delta g^{(0)}]  \right] .
\end{align} In these expressions, the presymplectic potentials of the boundary Einstein-Hilbert theory $\Theta_{EH}^a$ and the quadratic curvature gravity $\Theta^a_{QCG(1)}$ and $\Theta^a_{QCG(2)}$, appear naturally (see appendix \ref{New massive gravity}). Let us now make some general comments. 

\begin{itemize}[label=$\rhd$]
\item The infinitesimal charges \eqref{surface charge expression} are associated with the most generic Al(A)dS$_{d+1}$ spacetimes written in Starobinsky/Fefferman-Graham gauge \eqref{FG gauge}. They are finite even if the boundary metric is varied, as direct consequence of the holographic renormalization of the symplectic structure controlled by \eqref{renormalized presymplectic current}. This result generalizes previous considerations \cite{Compere:2008us} by allowing field-dependence of the parameters \cite{Barnich:2007bf,Adami:2020ugu}, both signs of the cosmological constant ($\eta = \pm 1$) and non-vanishing Weyl parameters $\sigma$. It includes all the previous analyses with more restrictive boundary conditions \cite{Brown:1986nw, Henneaux:1985tv , Henneaux:1985ey , Ashtekar:1999jx , Troessaert:2013fma , Compere:2013bya , Papadimitriou:2005ii, Perez:2016vqo , Aros:1999kt , Fischetti:2012rd , Alessio:2020ioh}. 

\item Furthermore, the infinitesimal charges \eqref{surface charge expression} are $1$-forms on the solution space, and are generically non-integrable. Obtaining finite charges requires both the prescription to select a preferred integrable part \cite{Wald:1999wa , Compere:2018ylh , Chandrasekaran:2020wwn} and the integration on a path in the solution space \cite{Barnich:2007bf}. In addition, the infinitesimal charges \eqref{surface charge expression} are generically not conserved (see equations \eqref{renormalized presymplectic current} and \eqref{fundamental formula}). The source of the non-conservation can be related to radiating degrees of freedom but also to Weyl anomalies. In particular, for $d=2$, the breaking in the conservation law was interpreted in \cite{Alessio:2020ioh} as an anomalous Ward-Takahashi identity for the Weyl symmetry in the dual theory.

\item Finally, an important observation is that the charges associated with the Weyl parameter $\sigma (x)$ vanish for odd $d$, but are generically non-vanishing for even $d$. Let us provide an interpretation of this phenomenon. The asymptotic symmetry group of a theory is defined as the quotient between residual gauge diffeomorphisms and trivial gauge diffeomorphisms. Here, a residual gauge diffeomorphism $\xi$ is trivial if \eqref{surface charge expression} vanishes, \textit{i.e.} $\ndelta H_\xi [g] = 0$. The action of the asymptotic symmetries modifies the state of the system, while the trivial gauge diffeomorphisms do not affect it and are pure redundancies of the theory. Therefore, in odd $d$, since the Weyl charges vanish, we are free to perform Weyl rescalings without affecting the physics. This corresponds to the freedom to choose the finite part of the conformal factor in the conformal compactification process. On the contrary, for even $d$, Weyl charges are non vanishing. Henceforth, Weyl rescalings are not pure redundancies of the theory and different conformal factors in the conformal compactification process lead to physically inequivalent situations. Of course, this statement is very natural from the holographic perspective because of the presence of Weyl anomalies in even $d$. Indeed, in this case, we are not free to perform Weyl rescalings on the induced boundary metric because of the Weyl anomaly, which is consistent with the bulk result. 
\end{itemize}

\subsection{Charge algebra}

The infinitesimal charge expression \eqref{surface charge expression} being non-integrable, one can not use the standard results of the representation theorem to derive the charge algebra \cite{Barnich:2001jy, Barnich:2007bf}. Instead, we show that, using the Barnich-Troessaert prescription for the modified bracket \cite{Barnich:2011mi}, we can still obtain a meaningful charge algebra. We choose the following split between integrable and non-integrable parts in \eqref{surface charge expression}: 
\begin{equation}
\ndelta H_\xi [g]  = \delta  H_\xi [g] + \Xi_\xi [g; \delta g] 
\label{total charge}
\end{equation} where
\begin{equation}
\begin{split}
H_\xi [g] &= \int_{S_\infty}  (\D^{d-1} x) \left[  \sqrt{|g^{(0)}|} g^{tc}_{(0)} T^{[d]}_{bc} \bar{\xi}^b \right], \\
\Xi_\xi  [g; \delta g] &= \int_{S_\infty}  (\D^{d-1} x) \left[ - \frac{1}{2} \sqrt{|g^{(0)}|} \bar{\xi}^t T^{bc}_{[d]} \delta g_{bc}^{(0)} + W^{[d]t}_\sigma [g; \delta g] \right] - H_{\delta \xi} [g] .
\end{split}
\label{split of the charge}
\end{equation} Note that this split is ambiguous since one can always shift the integrable part and the non-integrable part with a certain expression $\Delta H_\xi[g]$ as
\begin{equation}
H_\xi [g] \to H_\xi [g] + \Delta H_\xi [g] , \quad \Xi_\xi  [g; \delta g] \to \Xi_\xi  [g; \delta g] - \delta (\Delta H_\xi [g]) ,
\label{shift}
\end{equation} without affecting the total charge \eqref{total charge}.

Using the Barnich-Troessaert prescription for the modified bracket \cite{Barnich:2011mi}, 
\begin{equation}
\{ H_{\xi_1} [g] , H_{\xi_2} [g] \}_\star \equiv \delta_{\xi_2} H_{\xi_1}[g] + \Xi_{\xi_2} [g; \delta_{\xi_1} g] ,
\label{BT bracket}
\end{equation} we show in appendix \ref{Derivation of the charge algebra} that
\begin{equation}
\boxed{\{ H_{\xi_1} [g] , H_{\xi_2} [g] \}_\star = H_{[\xi_1, \xi_2]_\star}[g] + K^{[d]}_{\xi_1, \xi_2} [g]   }
\label{charge algebra}
\end{equation} where the bracket of vector fields $[\xi_1, \xi_2]_\star$ is given in \eqref{eq:VectorAlgebra}. The field-dependent $2$-cocycle $K^{[d]}_{\xi_1, \xi_2} [g]$ appearing in the right-hand side of \eqref{charge algebra} vanishes for odd $d$, \textit{i.e.} $K^{[2k+1]}_{\xi_1, \xi_2} [g]=0$ ($k\in\mathbb N_0$). For $d=2$, we have explicitly
\begin{equation}
\begin{split}
K_{\xi_1, \xi_2}^{[2]}[g] = \frac{\ell }{16 \pi G} \int_{S_\infty} (\D^{d-1} x) \sqrt{|g^{(0)}|} \Big[ 2\left(\sigma_1 D^t \sigma_2 - \sigma_2 D^t \sigma_1 \right)+R^{(0)} \left(\sigma_1 \bar\xi_2^t - \sigma_2 \bar\xi^t_1 \right) \Big]. 
\end{split} 
\label{cocycle 3d}
\end{equation} In section \ref{Dirichlet}, we show that \eqref{cocycle 3d} reproduces the Brown-Henneaux central extension in three dimensions \cite{Brown:1986nw}, indicating the presence of a holographic Weyl anomaly \cite{Henningson:1998gx}. For $d = 4$, we obtain 
\begin{equation}
\begin{split}
K_{\xi_1, \xi_2}^{[4]}[g] = \frac{\eta\,\ell^3}{16\pi G} \int_{S_\infty} (\D^{d-1} x) \sqrt{|g^{(0)}|}  &\left[ \left( R^{tb}_{(0)}-\frac{1}{2}R^{(0)}g^{tb}_{(0)}\right) \left(\sigma_1 D_b \sigma_2 - \sigma_2 D_b \sigma_1\right) ,\right. \\
&\quad\quad+ \frac{1}{4} \left. \left( R^{bc}_{(0)}R_{bc}^{(0)} - \frac{1}{3} R^2_{(0)}\right) \left( \sigma_1 \bar{\xi}^t_2 - \sigma_2 \bar{\xi}^t_1 \right)\right].
\end{split} \label{cocycle 5d}
\end{equation} The field-dependent $2$-cocycle is antisymmetric, $K_{\xi_1, \xi_2}^{[d]}[g] = - K_{\xi_2, \xi_1}^{[d]}[g]$, and satisfies the $2$-cocycle condition
\begin{equation}
K_{[\xi_1, \xi_2]_\star, \xi_3}^{[d]}[g] + \delta_{\xi_3} K_{\xi_1, \xi_2}^{[d]}[g] + \text{cyclic(1,2,3)} = 0 . 
\label{2 cocycle condition}
\end{equation} Some details of this computation can be found in appendix \ref{Derivation of the charge algebra}. The explicit form of the field-dependent $2$-coycle relies on the choice of split between integrable and non-integrable parts \eqref{split of the charge}. Indeed, under a shift \eqref{shift}, the field-dependent $2$-cocycle transforms as
\begin{equation}
K_{\xi_1, \xi_2}^{[d]}[g] \to K_{\xi_1, \xi_2}^{[d]}[g] + \delta_{\xi_2} (\Delta H_{\xi_1}[g]) - \delta_{\xi_1} (\Delta H_{\xi_2}[g]) - \Delta H_{[\xi_1, \xi_2]_\star}[g] . 
\end{equation} However, the structure of the algebra \eqref{charge algebra} is not affected by this shift. 

In addition to the mathematical consistencies that we have mentioned, the charge algebra \eqref{charge algebra} also contains all the information about the flux-balance laws associated with the charges $H_\xi[g]$. We refer to \cite{Barnich:2011mi, Barnich:2013axa,Barnich:2011ty,Compere:2019gft,Godazgar:2018vmm} for the details (see also \cite{Compere:2020lrt} for a similar discussion in the Al(A)dS$_4$ context).  

\section{Application to more restrictive boundary conditions}
\label{Application to more restrictive boundary conditions}

We now apply our general results to specific cases of boundary conditions that have been considered in previous analyses. This presentation does not pretend to be exhaustive. Instead, we show that the results of section \ref{sec:Charge algebra in asymptotically locally} reduce consistently to some well-known results of the literature. More specifically, we consider Dirichlet \cite{Brown:1986nw , Hawking:1983mx , Ashtekar:1984zz , Henneaux:1985tv , Henneaux:1985ey , Ashtekar:1999jx , Papadimitriou:2005ii} and Neumann \cite{Compere:2008us} boundary conditions in asymptotically AdS$_{d+1}$ spacetimes, and partial Dirichlet boundary conditions that lead to the $\Lambda$-BMS$_{d+1}$ group in asymptotically (A)dS$_{d+1}$ spacetimes \cite{Compere:2019bua , Compere:2020lrt}. The two first cases are examples of conservative boundary conditions (on-shell stationary action, integrable and conserved charges), while the third case illustrates the larger class of leaky boundary conditions that we are considering here.

\subsection{Dirichlet boundary conditions}
\label{Dirichlet}

\subsubsection{Variational principle}
\label{variational principle}

As explained in \cite{Papadimitriou:2005ii}, a natural boundary condition to impose in asymptotically AdS$_{d+1}$ spacetimes at infinity is to fix the conformal class of the boundary metric. More precisely, this amounts to impose that the boundary metric is fixed, up to a conformal factor 
\begin{equation}
\delta g^{(0)}_{ab} = \lambda(x^c) g^{(0)}_{ab} .
\label{conformal class fixed}
\end{equation}  Taking \eqref{conformal class fixed} into account, the general result of the variation of the on-shell action encoded in \eqref{Theta ren pullback} and \eqref{var principle theta} reduces to
\begin{equation}
\delta S_{ren} = \frac{1}{2} \int_{\mathscr{I}} \sqrt{- g^{(0)}} \lambda \mathcal{T}^{[d]}
\end{equation} which reproduces the integrated Weyl anomaly (this corresponds to equation (3.48) of \cite{Papadimitriou:2005ii}). As a consequence of \eqref{trace even}, the action is stationary on solutions when $d$ is odd. However, this is generically not true when $d$ is even and one has to pick up a specific representative so that $\delta S_{ren} = 0$ on-shell. This is the point of view adopted in \textit{e.g.} \cite{Brown:1986nw ,  Henneaux:1985tv , Henneaux:1985ey} where the leading order of the bulk metric is taken to be a specific boundary metric. Therefore, we freeze the boundary metric $g^{(0)}_{ab}$ on the phase space as
\begin{equation}
g^{(0)}_{ab} \D x^a \D x^b = -\frac{1}{\ell^2} \D t^2 + \mathring{q}_{AB} \D x^A \D x^B 
\label{Dirichlet boundary conditions}
\end{equation} where $\mathring{q}_{AB}$ is the unit $(d-1)$-sphere metric and $x^a = (t/\ell, x^A)$, $A=2, \ldots , d$. For $d=2$, the metric $\mathring{q}_{AB}$ has only one component that we take $\mathring{q}_{\phi \phi} = 1$. In \eqref{Dirichlet boundary conditions}, we assume that $t$, the time coordinate on the boundary cylinder, has unit of length, while the transverse coordinates $x^A$ are still dimensionless.

Taking \eqref{Dirichlet boundary conditions} into account, the renormalized presymplectic current \eqref{renormalized presymplectic current} vanishes at $\mathscr{I}_{\text{AdS}}$, \textit{i.e.}
\begin{equation}
\bm \omega_{ren}[g;\delta g, \delta g]\Big|_{\mathscr{I}_{\text{AdS}}} = 0  .
\label{Vanishing of the symp form}
\end{equation} By virtue of equation \eqref{fundamental formula}, this implies that there is no flux leaking through the spacetime boundary (the gravitational waves are bouncing on the spacetime boundary) and the Cauchy problem is well defined \cite{Ishibashi:2004wx}. Furthermore, as already mentioned, the action is stationary on solutions since $\bm \Theta_{ren}[g;\delta g] |_{\mathscr{I}_{\text{AdS}}} =0$ \cite{Papadimitriou:2005ii} (see equations \eqref{Theta ren pullback} and \eqref{var principle theta}). Notice that the boundary conditions \eqref{Dirichlet boundary conditions} are not relevant in asymptotically dS$_{d+1}$ spacetimes since they would strongly constrain the Cauchy problem by eliminating the solutions with radiation reaching the future spacelike boundary $\mathscr{I}^+_{\text{dS}}$ (see Figure \hyperlink{fig:dS}{1.(b)}).

\subsubsection{Asymptotic symmetry algebra}
\label{Asymptotic symmetry algebra Dirichlet}

The residual gauge diffeomorphisms \eqref{AKV 1} and \eqref{AKV 2} preserving the boundary conditions \eqref{Dirichlet boundary conditions} are constrained through $\delta_\xi g^{(0)}_{ab} = 0$. Using \eqref{eq:action solution space}, this yields
\begin{equation}
\mathcal{L}_{\bar{\xi}} g^{(0)}_{ab} = 2 \sigma g^{(0)}_{ab} , \quad \sigma = \frac{1}{d} D_c \bar{\xi}^c , 
\label{CKV eq}
\end{equation} meaning that the boundary diffeomorphisms $\bar{\xi}^a$ are conformal Killing vectors of $g^{(0)}_{ab}$. For $d> 2$, the equation \eqref{CKV eq} can be rewritten equivalently as
\begin{equation}
\begin{split}
&\partial_t \bar{\xi}^t = \frac{1}{(d-1)} D_A \bar{\xi}^A , \qquad \partial_t \bar{\xi}^A = \frac{1}{\ell^2} \mathring{q}^{AB} D_B \bar{\xi}^t , \qquad \sigma = \frac{1}{(d-1)} D_A \bar{\xi}^A, \\
&D_A \bar{\xi}_B + D_B \bar{\xi}_A = \frac{2}{(d-1)} D_C \bar{\xi}^C \mathring q_{AB}
\end{split}
\label{CKV eq decomposed}
\end{equation} where the last equation is the conformal Killing equation on the unit $(d-1)$-sphere metric. As discussed in \cite{Henneaux:1985tv , Henneaux:1985ey , Barnich:2013sxa , Compere:2019bua}, the asymptotic symmetry algebra formed by the residual gauge diffeomorphisms \eqref{AKV 1} and \eqref{AKV 2} satisfying \eqref{CKV eq decomposed} is the conformal algebra in $d$ dimensions, namely $SO(d, 2)$. Assuming the field-independence of the parameters $\bar{\xi}^a$, which is consistent with the constraints \eqref{CKV eq decomposed}, the algebra \eqref{eq:VectorAlgebra} reduces to
\begin{equation}
[\xi(\bar \xi_{1}^a),\xi(\bar{\xi}_{2}^a)]_\star =  \xi (\hat{\bar \xi}^a) , \qquad \hat{\bar{\xi}}^a = \bar{\xi}_{1}^b \partial_b  \bar{\xi}_{2}^a  - \bar{\xi}_{2}^b \partial_b  \bar{\xi}_{1}^a , \label{Dirichlet algebra xi}
\end{equation}
For $d=2$, the equation \eqref{CKV eq} infers
\begin{equation}
\partial_t \bar{\xi}^t = \partial_\phi \bar{\xi}^\phi , \qquad\partial_t \bar{\xi}^\phi = \frac{1}{\ell^2} \partial_\phi \bar{\xi}^t , \qquad \sigma =  \partial_\phi \bar{\xi}^\phi.
\label{Dirichlet 2d}
\end{equation} Performing the coordinate transformation $x^\pm = \frac{t}{\ell} \pm \phi$ and expressing the parameters $\bar{\xi}^t$ and $\bar{\xi}^\phi$ as 
\begin{equation}
\bar{\xi}^t = \frac{\ell}{2} (Y^+ + Y^- ), \qquad \bar{\xi}^\phi = \frac{1}{2} (Y^+ - Y^-),
\label{redefinition generators}
\end{equation} equation \eqref{Dirichlet 2d} implies $Y^\pm = Y^\pm (x^\pm)$ \cite{Barnich:2012aw}, and \eqref{Dirichlet algebra xi} becomes
\begin{equation}
[\xi(Y^\pm),\xi(Y^\pm)]_\star =  \xi (\hat{Y}^\pm) , \qquad \hat{Y}^\pm = Y_1^\pm \partial_\pm Y_2^\pm - Y_2^\pm \partial_\pm Y_1^\pm.
\label{vector algebra 2d}
\end{equation} Finally, expanding the parameters in modes as $Y^\pm = \sum_{m \in \mathbb{Z}} Y_m^\pm  l_m^\pm$, with $l_m^\pm = e^{i m x^\pm}$, the commutation relations \eqref{vector algebra 2d} yield
\begin{equation}
i [ l_m^\pm , l_n^\pm ] = (m-n) l_{m+n}^\pm  , \qquad [ l_m^\pm , l_n^\mp ] = 0  ,
\end{equation} which corresponds to the double copy of the Witt algebra, nameley Diff$(S^1)$ $\oplus$ Diff$(S^1)$ \cite{Brown:1986nw}.

\subsubsection{Charge algebra}
\label{charge algebraaa diri}

Inserting \eqref{Dirichlet boundary conditions} into \eqref{surface charge expression}, we deduce that the infinitesimal charges associated with Dirichlet boundary conditions are integrable, \textit{i.e.} $\ndelta H_\xi [g] = \delta H_\xi [g]$ with
\begin{equation}
H_\xi [g] =  \frac{1}{\ell}\int_{S_\infty}  (\D^{d-1} x) \sqrt{\mathring{q}}\left( {T^t}_b \bar{\xi}^b \right) 
\label{Bare Dirichlet charges}
\end{equation} 
and $\mathring{q} = \det (\mathring{q}_{AB})$. This corresponds to the Noether charge of a conformal field theory obtained by contracting the stress-energy tensor with a conformal Killing vector. Now integrating on a path in the solution space and requiring that the charges vanish for global AdS$_{d+1}$, we obtain
\begin{equation}
\tilde H_\xi [g] = \frac{1}{\ell}\int_{S_\infty}  (\D^{d-1} x) \sqrt{\mathring{q}}\left( {T^t}_b \bar{\xi}^b \right) - N_\xi , \qquad N_\xi \equiv H_\xi [g]\Big|_{\text{AdS}}.
\label{Dirichlet charges}
\end{equation}
Here $N_\xi$ denotes \eqref{Bare Dirichlet charges} evaluated for global AdS$_{d+1}$. As a consequence of \eqref{fundamental formula} and \eqref{Vanishing of the symp form}, the charges \eqref{Dirichlet charges} are conserved in time. 

Now, we bring the boundary conditions \eqref{Dirichlet boundary conditions} at the level of the charge algebra \eqref{charge algebra}. Since the charges \eqref{Dirichlet charges} are integrable, the standard results of the representation theorem \cite{Barnich:2001jy, Barnich:2007bf} are recovered. Indeed, the Barnich-Troessaert bracket \eqref{BT bracket} reduces to the standard Peierls bracket for integrable charges \cite{Peierls:1952cb , Harlow:2019yfa} ($\Xi_{\xi_2}[g; \delta_{\xi_1} g ] =0$), namely
\begin{equation}
\{ H_{\xi_1} [g], H_{\xi_2} [g] \} = \delta_{\xi_2} H_{\xi_1} [g]  .
\end{equation} Henceforth, the charge algebra \eqref{charge algebra} yields
\begin{equation}
\{ \tilde H_{\xi_1} [g], \tilde H_{\xi_2} [g] \} = \tilde H_{[\xi_1, \xi_2]_\star}[g] + \tilde K_{\xi_1, \xi_2}^{[d]}, \qquad \tilde K_{\xi_1, \xi_2}^{[d]} \equiv K_{\xi_1, \xi_2}^{[d]} + N_{[\xi_1,\xi_2]_\star}
\label{charge algebra 2d}
\end{equation} where $\tilde K_{\xi_1, \xi_2}^{[d]}$ vanishes for odd $d$, \textit{i.e.} $\tilde K_{\xi_1, \xi_2}^{[2k+1]} = 0$ ($k \in \mathbb{N}_0$). In odd spacetime dimensions (even $d$), taking \eqref{Dirichlet boundary conditions} into account, the $2$-cocycle $\tilde K_{\xi_1, \xi_2}^{[d]}$ is field-independent and becomes a central extension that satisfies the standard $2$-cocycle condition 
\begin{equation}
\tilde K_{[\xi_1, \xi_2]_\star, \xi_3}^{[d]} + \text{cyclic(1,2,3)} = 0 
\end{equation} as a direct consequence of \eqref{2 cocycle condition} and \eqref{Dirichlet boundary conditions}. 

In particular, for $d=2$, the $2$-cocycle reduces to the Brown-Henneaux central extension \cite{Brown:1986nw}. Indeed, inserting \eqref{Dirichlet boundary conditions} and \eqref{Dirichlet 2d} into \eqref{cocycle 3d}, and adding the contribution of the global AdS$_3$ background as in \eqref{charge algebra 2d}, we readily obtain 
\begin{equation}
\tilde K^{[2]}_{\xi_1, \xi_2} = -\frac{1}{8\pi G} \int_0^{2\pi} \D \phi \Big[\partial_\phi \bar \xi_1^\phi \partial_\phi^2 \bar \xi^t_2 - \frac{1}{2}\bar\xi^t_1 \partial_\phi \bar \xi_2^\phi -\frac{1}{2} \bar \xi^\phi_1 \partial_\phi \bar \xi^t_2 - (1\leftrightarrow 2) \Big].
\label{intermediate step}
\end{equation} 
Integrating by parts and throwing away the total $\phi$ derivatives, the central extension \eqref{intermediate step} can be expressed in terms of the parameters $Y^+$ and $Y^-$ defined in \eqref{redefinition generators} as 
\begin{equation}
\tilde K^{[2]}_{\xi_1, \xi_2} = \frac{\ell}{16\pi G} \int_0^{2\pi} \D \phi \,  \Big[ Y^+_1 (\partial_+^3 Y^+_2 + \partial_+ Y^+_2) + Y^-_1 (\partial^3_- Y^-_2 + \partial_- Y^-_2) \Big] .
\end{equation} Finally, writing $\texttt L_m^\pm = \tilde H_{\xi(l^\pm_m)}[g]$ in \eqref{charge algebra 2d}, we recover the double copy of the Virasoro algebra
\begin{equation}
i \{\texttt  L_m^\pm ,\texttt  L_n^\pm \} = (m-n) \texttt L_{m+n}^\pm - \frac{c^\pm}{12} m (m^2-1) \delta^0_{m+n} , \qquad \{ \texttt L_m^\pm , \texttt L_n^\mp \} = 0 
\end{equation} where 
\begin{equation}
\boxed{
c^\pm = \frac{3\ell}{2G},
}
\end{equation}
which corresponds to the results of \cite{Brown:1986nw}.

For $d=4$, the $2$-cocycle $\tilde K_{\xi_1,\xi_2}^{[4]}$ vanishes \cite{Henneaux:1985ey}. In fact, inserting \eqref{Dirichlet boundary conditions} and \eqref{CKV eq decomposed} into \eqref{cocycle 5d}, and adding the contribution of the global AdS$_5$ background as in \eqref{charge algebra 2d}, one readily finds
\begin{equation}
\tilde K^{[4]}_{\xi_1, \xi_2} = \frac{\ell^2}{48 \pi G} \int_{S_\infty} (\text{d}^{d-1}x) \, \sqrt{\mathring q} \left[D_A \bar\xi^A_1 D_B D^B \bar\xi_2^t - (1\leftrightarrow 2)\right] - \frac{3\ell^2}{64 \pi G}  \int_{S_\infty} (\text{d}^{d-1}x) \, \sqrt{\mathring q}\, \hat{\bar{\xi}}^t .
\end{equation} Integrating by parts and using $D_B D^B (D_A \bar\xi^A_1) = -3 D_A \bar\xi^A_1$, which is a consequence of the conformal Killing equation \eqref{CKV eq decomposed}, we get
\begin{equation}
\begin{split}
\tilde K^{[4]}_{\xi_1, \xi_2} &= -\frac{\ell^2}{16 \pi G} \int_{S_\infty} (\text{d}^{d-1}x) \, \sqrt{\mathring q} \left[D_A \bar\xi^A_1 \bar\xi_2^t - (1\leftrightarrow 2)\right] - \frac{3\ell^2}{64 \pi G}  \int_{S_\infty} (\text{d}^{d-1}x) \, \sqrt{\mathring q}\, \hat{\bar{\xi}}^t = 0. \label{cocycle 4 zero}
\end{split} 
\end{equation} To obtain the second equality, we have integrated by parts and used the $t$-component of the commutation relations \eqref{Dirichlet algebra xi}, namely $\hat{\bar{\xi}}^t = \bar{\xi}_{1}^A D_A \bar{\xi}_{2}^t  + \frac{1}{3} \bar{\xi}_{1}^t D_A \bar{\xi}_{2}^A   - (1 \leftrightarrow 2 )$. From \eqref{cocycle 4 zero}, we see that the $2$-cocycle \eqref{cocycle 5d} with Dirichlet boundary conditions satisfies $K^{[4]}_{\xi_1, \xi_2} = - N_{[\xi_1, \xi_2]_\star}$. This means that this $2$-cocycle is a coboundary that is reabsorbed by adjusting the zero of the charges as in \eqref{Dirichlet charges}, see \textit{e.g.} \cite{Barnich:2007bf ,Barnich:2001jy , Compere:2018aar}. As discussed in \cite{Henneaux:1985ey}, one can actually show that the $2$-cocycle $\tilde K^{[d]}_{\xi_1, \xi_2}$ appearing in \eqref{charge algebra 2d} vanishes for any $d> 2$.

%

\subsection{Neumann boundary conditions}

\subsubsection{Residual diffeomorphisms}

In the previous section, we mentioned that the Dirichlet boundary conditions \eqref{Dirichlet boundary conditions} lead to a vanishing presymplectic current at $\mathscr{I}_{\text{AdS}}$ (see equation \eqref{Vanishing of the symp form}) by freezing the boundary metric on the phase space, namely $\delta g^{(0)}_{ab} = 0$. Alternatively, one could also consider an other set of boundary conditions in asymptotically AdS$_{d+1}$ spacetimes that yields \eqref{Vanishing of the symp form} by requiring $\delta T_{ab}^{[d]} = 0$ and keeping the boundary metric $g^{(0)}_{ab}$ free. These are called Neumann boundary conditions. From \eqref{renormalized presymplectic current}, we readily see that \eqref{Vanishing of the symp form} holds and therefore, the Cauchy problem is well posed \cite{Ishibashi:2004wx}. As discussed \textit{e.g.} in \cite{Papadimitriou:2005ii}, an on-shell stationnary action can be obtained (from \eqref{Theta ren pullback} and \eqref{var principle theta}, this is automatic for odd $d$, while for even $d$, a finite boundary Lagrangian term proportional to the Weyl anomaly has to be added to the action). Following \cite{Compere:2008us}, we impose the stronger condition
\begin{equation}
T_{ab}^{[d]} = 0 .
\label{Neumann}
\end{equation} The latter condition allows to derive clearer constraints on the residual gauge diffeomorphisms. We repeat briefly the discussion of \cite{Compere:2008us} to apply our general framework. 

In the odd $d$ case, the holographic stress-energy tensor \eqref{holographic stress-energy tensor} transforms homogeneously under the residual gauge diffeomorphisms \eqref{AKV 1} and \eqref{AKV 2} (see equation \eqref{eq:action solution space 2}). Hence, the boundary condition \eqref{Neumann} does not imply any constraint on the parameters $\sigma$ and $\bar{\xi}^a$. However, in the even $d$ case, the transformation of the holographic stress-energy tensor involves inhomogeneous terms in $A_{ab}^{[d]}[\sigma]$. Henceforth, one has to impose $\sigma = 0$ for \eqref{Neumann} to be satisfied in even $d$. 

\subsubsection{Charge algebra}

In odd $d$, inserting the condition \eqref{Neumann} into \eqref{surface charge expression} readily yields $\ndelta H_\xi [g] = 0$. Therefore, the residual gauge diffeomorphisms \eqref{AKV 1} and \eqref{AKV 2} are trivially represented. Similarly, in even $d$, inserting the condition \eqref{Neumann} into \eqref{surface charge expression} and taking into account that $\sigma = 0$, we obtain that the charges are zero. The asymptotic symmetry group associated to Neumann boundary conditions is therefore trivial. From the point of view of the dual theory, the boundary diffeomorphisms are pure gauge transformations, which indicates the presence of quantum gravity on the boundary. The latter is Weyl invariant for odd $d$.

\subsection{Leaky boundary conditions and \texorpdfstring{$\Lambda$-BMS$_{d+1}$}{Lambda-BMS d+1}}

\subsubsection{Boundary conditions and asymptotic symmetry algebra}
\label{Boundary conditions and asymptotic symmetry algebra LBMS}

Instead of imposing the Dirichlet boundary conditions \eqref{Dirichlet boundary conditions} that completely freeze the boundary metric $g^{(0)}_{ab}$, we require weaker constraints that allow for some fluctuations of the transverse metric components: 
\begin{equation}
g_{tt}^{(0)} = -\frac{\eta}{\ell^2} , \qquad g^{(0)}_{tA} = 0 , \qquad \sqrt{|g^{(0)}|} = \frac{1}{\ell} \sqrt{\mathring{q}} 
\label{BC Lambda BMS}
\end{equation} where $\mathring{q}$ is a fixed volume of a codimension 2 surface taken to be the determinant of the unit round $(d-1)$-sphere metric $\mathring{q}_{AB}$ (for $d=2$, we take $\mathring{q} = 1$ on $S^1$) in order to include the global (A)dS$_{d+1}$ spacetime in the phase space. These boundary conditions are a direct $(d+1)$-dimensional generalization of the boundary conditions proposed in \cite{Compere:2019bua} in the four-dimensional case. They are relevant for both signs of the cosmological constant. Indeed, the boundary gauge fixing \eqref{BC Lambda BMS} is always reachable using the freedom that we have on the residual gauge diffeomorphisms \eqref{AKV 1} and \eqref{AKV 2} \cite{Compere:2019bua}. Therefore, it does not constrain the Cauchy problem in asymptotically dS$_{d+1}$ spacetimes and allows for some flux at the boundary (see section \ref{Lambda BMS charge algebra section}). Fundamentally, it amounts to introduce an additional background structure to the Starobinsky/Fefferman-Graham foliation, consisting of a boundary foliation of constant $t$ codimension 2 hypersurfaces on which the volume form $\sqrt{\mathring{q}}(\D^{d-1}x)$ is fixed ($\delta \sqrt{\mathring{q}}=0$).

Requiring these boundary conditions to be preserved under the residual gauge diffeomorphisms generated by \eqref{AKV 1} and \eqref{AKV 2} yields the following conditions on the parameters:
\begin{equation}
\partial_t \bar{\xi}^t = \frac{1}{(d-1)} D_A \bar{\xi}^A , \qquad \partial_t \bar{\xi}^A = \frac{\eta}{\ell^2} g^{AB}_{(0)} D_B \bar{\xi}^t , \qquad \sigma = \frac{1}{(d-1)} D_A \bar{\xi}^A .
\label{equation Lambda BMS}
\end{equation} Since these equations involves explicitly the transverse metric $g^{AB}_{(0)}$, which is a dynamical field of the theory, the parameters $\bar{\xi}^t$ and $\bar{\xi}^A$ are field-dependent. Taking \eqref{equation Lambda BMS} into account, the residual gauge diffeomorphisms satisfy the following commutation relations with the modified Lie bracket \eqref{modified bracket}:
\begin{equation}
[{\xi}(\bar{\xi}_{1}^t,\bar{\xi}_{1}^A ),{\xi}(\bar{\xi}_{2}^t ,\bar{\xi}_{2}^A )]_\star =  \xi (\hat{\bar{\xi}}_{1}^t,\hat{\bar{\xi}}_{1}^A) \label{eq:VectorAlgebra1Lambda}
\end{equation}
where
\begin{equation}
\begin{split}
\hat{\bar{\xi}}^t &= \bar{\xi}_{1}^A D_A \bar{\xi}_{2}^t  + \frac{1}{(d-1)} \bar{\xi}_{1}^t D_A \bar{\xi}_{2}^A   - \delta_{\xi_1} \bar{\xi}_{2}^t - (1 \leftrightarrow 2 ), \\
\hat{\bar{\xi}}^A &=  \bar{\xi}^B_{1} D_B \bar{\xi}^A_{2} + \frac{\eta}{\ell^2} \bar{\xi}^t_{1} g^{AB}_{(0)} D_B \bar{\xi}_{2}^t   - \delta_{\xi_1} \bar{\xi}_{2}^A - (1 \leftrightarrow 2).
\end{split} \label{eq:VectorAlgebraLambda}
\end{equation} This is a corollary of \eqref{eq:VectorAlgebra}. These commutation relations are field-dependent for generic $d$ and therefore depend on the point of the solution space we are looking at. Hence, this algebra of asymptotic symmetries constitutes rather a Lie algebroid \cite{Crainic , Barnich:2010xq , Barnich:2017ubf} that we call $\Lambda$-BMS$_{d+1}$. Let us mention some properties of it:
\begin{itemize}[label=$\rhd$]
\item At each point of the solution space, $\Lambda$-BMS$_{d+1}$ forms an infinite-dimensional algebra. Indeed, it always contains the infinite-dimensional algebra of area-preserving diffeomorphisms on the $(d-1)$-sphere ($\bar{\xi}^t = 0$, $\partial_t \bar{\xi}^A = 0$, $D_A \bar{\xi}^A = 0$) as a subalgebra \cite{Compere:2019bua}. 

\item In the flat limit $\ell \to \infty$ ($\Lambda \to 0$), $\Lambda$-BMS$_{d+1}$ reduces to the asymptotic symmetry algebra of asymptotically (locally) flat spacetimes, namely the (generalized) Bondi-Metzner-Sachs-van der Burg algebra in $d+1$ dimensions, written BMS$_{d+1}$ \cite{1962RSPSA.269...21B , Sachs:1962wk , Barnich:2010eb , Campiglia:2014yka , Campiglia:2015yka , Compere:2018ylh , Colferai:2020rte , Kapec:2015vwa, Campoleoni:2017qot , Pate:2017fgt ,  Aggarwal:2018ilg , Barnich:2006av , Hollands:2016oma , Herfray:2020rvq}. Indeed, taking $\ell \to \infty$ in \eqref{equation Lambda BMS}, one can find the explicit solutions for $\bar{\xi}^t$ and $\bar{\xi}^A$ given by
\begin{equation}
\bar{\xi}^t = T + \frac{t}{2} D_A Y^A, \quad \bar{\xi}^A = Y^A 
\label{solution BMS}
\end{equation} where $T=T(x^A)$ are the supertranslation parameters and $Y^A = Y^A (x^B)$ are the super-Lorentz parameters. At the level of the commutation relations, the flat limit gives $\delta_{\xi} \bar{\xi}^t = 0$, $\delta_\xi \bar{\xi}^A = 0$ and
\begin{equation}
\begin{split}
\hat{\bar{\xi}}^t &= \bar{\xi}_{1}^A D_A \bar{\xi}_{2}^t  + \frac{1}{(d-1)} \bar{\xi}_{1}^t D_A \bar{\xi}_{2}^A   - (1 \leftrightarrow 2 ), \\
\hat{\bar{\xi}}^A &=  \bar{\xi}^B_{1} D_B \bar{\xi}^A_{2}  - (1 \leftrightarrow 2), 
\end{split} \label{eq:VectorAlgebraBMS4}
\end{equation} or, taking \eqref{solution BMS} into account,
\begin{equation}
\begin{split}
\hat T &= Y_1^A D_A T_2  + \frac{1}{(d-1)} T_1 D_A Y^A_2   - (1 \leftrightarrow 2 ), \\
\hat Y^A &=  Y^B_1 D_B Y^A_2  - (1 \leftrightarrow 2).
\end{split}
\end{equation} These commutation relations precisely correspond to those of the generalized BMS$_n$ algebra given by Supertranslations $\loplus$ Diff$(S^{d-1})$. 

\item For $d= 2$, the codimension 2 boundary metric $g_{AB}^{(0)}$ with fixed determinant $\mathring{q} = 1$ has only one component $g^{(0)}_{\phi\phi} = 1$ and the boundary conditions \eqref{BC Lambda BMS} reduces to the Dirichlet boundary conditions \eqref{Dirichlet boundary conditions}. Henceforth, the $\Lambda$-BMS$_3$ algebroid is an algebra and corresponds to the infinite-dimensional conformal algebra in two dimensions Diff$(S^1)$ $\oplus$ Diff$(S^1)$ discussed in section \ref{Asymptotic symmetry algebra Dirichlet}. 

\item For $d = 3$, the structure of $\Lambda$-BMS$_4$ was investigated in details in \cite{Compere:2019bua , Compere:2020lrt}. In particular, some explicit solutions for the generators \eqref{equation Lambda BMS} were obtained using the decomposition of $\bar{\xi}^A$ into a curl-free and a divergence-free part available on the $2$-sphere. 

\item When the codimension 2 boundary metric $g^{(0)}_{AB}$ is frozen to be the $(d-1)$-sphere metric, $g^{(0)}_{AB} = \mathring{q}_{AB}$ (for $d> 2$), we recover the Dirichlet boundary conditions \eqref{Dirichlet boundary conditions}. In this case, $\Lambda$-BMS$_{d+1}$ reduces to $SO(d+1,1)$ for $\eta = -1$ and $SO(d,2)$ for $\eta = +1$. In the flat limit $\ell \to \infty$, these algebras reduce to the Poincaré algebra (see also the discussion in section \ref{sec:Comments}). 

\end{itemize}

\subsubsection{$\Lambda$-BMS$_{d+1}$ charge algebra}
\label{Lambda BMS charge algebra section}

Taking the boundary conditions \eqref{BC Lambda BMS} into account, the renormalized presymplectic potential \eqref{Theta ren pullback} reduces to
\begin{equation}
\bm\Theta_{ren}[g;\delta g]\Big|_{\mathscr I} = - \frac{\sqrt{\mathring{q}}}{2 \ell} T^{AB}_{TF} \delta g_{AB}^{(0)} \, (\D^d x)
\end{equation} where $T^{AB}_{TF} = T^{AB} - \frac{1}{d-1} g^{AB}_{(0)} T^C_C$ is the trace-free part of the $(d-1)$-dimensional tensor $T^{AB}$. The associated presymplectic current is given by
\begin{equation}
\bm \omega_{ren}[g; \delta_1 g, \delta_2 g]\Big|_{\mathscr I} = - \frac{\sqrt{\mathring{q}}}{2 \ell} \delta_1 T^{AB}_{TF} \delta_2 g_{AB}^{(0)} \,(\D^d x) - (1 \leftrightarrow 2)  .
\label{non vanishing flux}
\end{equation} This implies that the boundary conditions \eqref{BC Lambda BMS} are leaky for $d>2$, \textit{i.e.} there is some flux going through the spacetime boundary. In the asymptotically dS$_{d+1}$ case, the radiation going through $\mathscr{I}^+_{\text{dS}}$ is expected since, as explained in section \ref{Boundary conditions and asymptotic symmetry algebra LBMS}, the boundary conditions \eqref{BC Lambda BMS} do not restrict the Cauchy problem. In the asymptotically AdS$_{d+1}$ case, the presence of a non-vanishing flux \eqref{non vanishing flux} through $\mathscr{I}_{\text{AdS}}$ yields a non-globally hyperbolic spacetime \cite{Ishibashi:2004wx,Barroso:2019cwp}. In the holographic perspective, this translates into the fact that the dual theory couples to an external system \cite{Giddings:2020usy , Jana:2020vyx , Compere:2008us , Troessaert:2015nia , Almheiri:2014cka , Maldacena:2016upp , Apolo:2014tua}.

Similarly, the $\Lambda$-BMS$_{d+1}$ charges are obtained by inserting the boundary conditions \eqref{BC Lambda BMS} into the general charge expressions \eqref{surface charge expression}. We have explicitly
\begin{equation}
\ndelta H_\xi [g] = \delta H_\xi [g] + \Xi_\xi [g;\delta g]  
\end{equation} where
\begin{equation}
\begin{split}
H_\xi [g] &= - \eta \ell \int_{S_\infty} (\D^{d-1} x) \sqrt{\mathring{q}} \left[  {T}_{tt} \bar{\xi}^t + T_{tB} \bar{\xi}^B \right], \\
\Xi_\xi  [g; \delta g] &= \int_{S_\infty} (\D^{d-1} x) \left[ - \frac{1}{2\ell} \sqrt{\mathring{q}}\, \bar{\xi}^t \, T^{BC}_{TF} \delta g_{BC}^{(0)} +W^{[d]}[g; \delta g] \right] - H_{\delta \xi} [g] .
\end{split}
\label{split of the charge Lambda BMS}
\end{equation} The term $W^{[d]}[g;\hspace{-2pt} \delta g]$ depends on the dimension and is obtained by inserting \eqref{BC Lambda BMS} into $W^{[d]t}_\sigma [g;\hspace{-2pt} \delta g]$. In particular, this contribution vanishes when $d$ is odd. For $d=2$, the conditions \eqref{BC Lambda BMS} become the Dirichlet boundary conditions and so $W^{[2]} = 0$. Finally, for $d=4$, taking into account that \eqref{BC Lambda BMS} holds, one obtains 
\begin{equation}
\begin{split}
R^{(0)}_{tt} &= -\partial_t l - l^{AB}l_{AB}\ , \quad R^{(0)}_{tA} = D^B l_{AB} - \partial_A l  , \\
R^{(0)}_{AB} &= R_{AB}[q] + \eta\,\ell^2 \left( (\partial_t+l)l_{AB} - 2 {l_A}^C l_{BC} \right) , \\
R^{(0)} &= R[q] + \eta\,\ell^2 \left( (\partial_t+l)l + l^{AB}l_{AB} \right) 
\end{split}
\end{equation}
where $q_{AB}\equiv g_{AB}^{(0)}$, $l_A^B \equiv \frac{1}{2} q^{AC}\partial_t q_{BC}$ and $l = l^A_A = 0$. We have
\begin{align}
W^{[4]}[g^{(0)};\delta g^{(0)}] &= \int_{S_\infty} (\D^{d-1}x) \,\sqrt{\mathring q}\, \Big( 4 \bar\xi^t M_{\sigma}[g^{(0)};\delta g^{(0)}] + 2 \bar\xi^A \partial_A N_\sigma[g^{(0)};\delta g^{(0)}] \Big)  , \nonumber \\
M_{\sigma}[g^{(0)};\delta g^{(0)}] &= \frac{1}{768\pi G} D_A D^A \Big[ R_{BC}[q]\delta q^{BC} + \eta\ell^2 (\partial_t l_{BC}-2 {l_A}^C l_{BC})\delta q^{BC} \Big]  , \nonumber \\
N_{\sigma}[g^{(0)};\delta g^{(0)}] &= \frac{\eta\ell^3}{16\pi G} \Big[ \frac{1}{12} \eta\ell D_B(D^D l_{CD}\delta q^{BC}) + \frac{\ell^4}{12} R^{BC}[q]\delta l_{BC} + \frac{\eta\ell^6}{24}(\partial_t l^{BC}+2l^{BD} {l_D}^C)\delta l_{BC} \nonumber \\
&\phantom{= \frac{\eta\ell^3}{16\pi G} \Big[} - \frac{\ell^4}{12} D_C D^D l_{BD}\delta q^{BC} + \frac{\ell^4}{24}\partial_t R_{BC}[q]\delta q^{BC} \nonumber \\
&\phantom{= \frac{\eta\ell^3}{16\pi G} \Big[} + \frac{\ell^4}{24} (\partial_t^2 l_{BC} - 2 {l_B}^D \partial_t l_{CD} + 4 {l_B}^D {l_D}^E l_{EC} )\delta q^{BC} \nonumber \\
&\phantom{= \frac{\eta\ell^3}{16\pi G} \Big[} + \frac{\ell^4}{36} (R[q] + \eta\ell^2)l_{DE}l^{DE})l_{BC}\delta q^{BC}\Big]  \label{relations d4}
\end{align} 
after performing several partial integrations on the boundary $3$-sphere and employing \eqref{equation Lambda BMS} explicitly.

As a corollary of \eqref{charge algebra}, one obtains the $\Lambda$-BMS$_{d+1}$ charge algebra. For $d=2k+1$ ($k \in\mathbb N_0$), the algebra closes without central extensions. The case $d=2$ has already been discussed in section \ref{charge algebraaa diri}. Finally, for $d=4$, using \eqref{relations d4}, we obtain the following expression for the field-dependent $2$-cocycle:
\begin{align}
K^{[4]}_{\xi_1,\xi_2}[g^{(0)}] &= \frac{\eta\,\ell^2}{16\pi G} \int (\D^{d-1}x)\,\sqrt{\mathring q} \, \times \ldots \nonumber \\
\ldots \times &\left\lbrace \frac{1}{18}(R[q]-\eta\ell^2 l_{AB}l^{AB})D_C \bar\xi^C_1 D_D D^D \bar\xi^t_2 - \frac{\eta\ell^2}{9}D_B l^{AB} D_C \bar\xi_1^C D_A D_D \bar\xi_2^D \right. \nonumber \\
&\quad \Big[ - \frac{\eta\ell^2}{12} D_B l^{AB} D^C l_{BC} + \frac{1}{12} R^{AB}[q]R_{AB}[q]  + \frac{\eta\ell^2}{6}R^{AB}[q](\partial_t l_{AB}-2 {l_A}^C l_{BC})  \nonumber \\
&\quad + \frac{\ell^4}{12}\partial_t l_{AB}\partial_t l^{AB} + \frac{\eta\ell^2}{12}\partial_t l_{AB} l^{AC}{l_C}^B - \frac{\eta\ell^2}{6}l^{AC}{l_C}^B\partial_t l_{AB}  + {l_A}^C {l_C}^B {l_D}^A {l_B}^D \nonumber \\
&\quad \left. - \frac{1}{36}R[q]^2 - \frac{\eta\ell^2}{18} l^{AB}l_{AB} + \frac{\ell^4}{18}(l_{AB}l^{AB})^2 \Big] D_E \bar\xi^E_1 \bar\xi^t_2 - (1\leftrightarrow 2) \right\rbrace .
\end{align} As discussed in \cite{Compere:2020lrt}, the form of the $\Lambda$-BMS$_4$ charge algebra may be affected by the addition of corner terms \cite{Horowitz:2019dym,Freidel:2020xyx, Freidel:2020svx , Freidel:2020ayo} at the level of the variational principle \eqref{ren action principle} that are necessary to take the flat limit at the level of the symplectic structure.

\section{Comments}
\label{sec:Comments}

We would like to end the discussion by briefly summarizing the main results obtained in the text and suggesting some perspectives for future analyses. In this article, we have proposed a general framework to analyse asymptotically locally AdS$_{d+1}$ spacetimes in Starobinsky/Fefferman-Graham gauge, without assuming any boundary condition other than the minimal falloffs allowing for conformal compactification. More specifically, after a renormalization procedure to remove the divergences from the symplectic structure, we have derived the infinitesimal charges and shown that they are generically neither integrable nor conserved. While the charges associated with boundary diffeomorphisms are non-vanishing in any dimension, we have shown that the Weyl charges are non-vanishing in odd spacetime dimensions only. This fact was interpreted from the point of view of the dual holographic theory, where the presence of Weyl anomalies does not allow to perform freely Weyl rescalings on the induced boundary metric. Using the Barnich-Troessaert modified bracket dealing with non-integrable expressions, we have derived the charge algebra and shown that it involves a field-dependent $2$-cocycle in odd spacetime dimensions. After this general analysis, we have specified our results to three set of boundary conditions: Dirichlet, Neumann and leaky boundary conditions. In particular, for asymptotically AdS$_3$ boundary conditions, we have shown that the $2$-cocycle of the charge algebra reduces consistently to the Brown-Henneaux central charge\footnote{Interestingly, when considering the new proposal of boundary conditions in AdS$_3$ discussed in \cite{Alessio:2020ioh}, the $2$-cocycle \eqref{cocycle 3d} also reproduces the central extension appearing in the abelian Weyl sector.}. Finally, considering the leaky boundary conditions has allowed us to define the $\Lambda$-BMS$_{d+1}$ Lie algebroid in any dimension. The latter reduces to the generalized BMS$_{d+1}$ symmetry algebra in the flat limit. 

As already emphasized in the introduction, the need for considering leaky boundary conditions or allowing for some fluctuations of the boundary structure is appealing. In asymptotically AdS spacetimes, we have seen that the flux arising through the spacetime boundary was present when coupling the holographic stress energy tensor to the boundary metric \eqref{Theta ren pullback}. The latter plays the role of source \cite{Compere:2008us , Troessaert:2015nia} and yields non-equilibrium physics and interactions between the system and the environment (see Figure \hyperlink{fig:AdS}{1.(a)}). From these considerations, it is not astonishing that the charges become non-integrable since we are facing with a dissipative system. Turning off the sources (which translates by freezing the boundary metric in Dirichlet boundary conditions \eqref{Dirichlet boundary conditions}) eliminates the leaks and the system is considered as isolated. The charges are then integrable and look like \eqref{Dirichlet charges}. It is worth noticing that starting with the boundary gauge fixing \eqref{BC Lambda BMS}, the requirement that the flux is frozen translates into $\delta g_{AB}^{(0)}=0$, which leads the symmetry breaking

\begin{equation}
\begin{aligned}
\begin{tikzpicture}
\coordinate (A) at (0,0);
\coordinate (B) at (3,0);
\draw (A) node[left]{\normalsize $\Lambda\text{-BMS}_{d+1}$};
\draw[-latex] (0,0) -- (B);
\draw (B) node[right]{\normalsize $SO(d,2)$};
\draw ($(A)!0.5!(B)$) node[above]{\scriptsize freezing sources};
\end{tikzpicture}
\label{symmetry breaking AdS}
\end{aligned}
\end{equation}
($d>2$), \textit{i.e.} the infinite-dimensional Lie algebroid $\Lambda\text{-BMS}_{d+1}$ reduces to the finite-dimensional symmetry algebra $SO(d,2)$ of the AdS$_{d+1}$ vacuum.

As we have discussed in the text, allowing some flux and fluctuating boundary metric in asympotically dS spacetimes is a natural requirement to keep interesting solutions in the analysis. Similarly, in the context of asymptotically flat spacetimes, letting flux going through null infinity is a necessary condition to consider radiating solutions. In this case, the asymptotic shear $C_{AB}$ plays the role of source and the charges are non-integrable \cite{Barnich:2011mi , Barnich:2013axa , Barnich:2019vzx,Herfray:2020rvq}. When asymptotically locally flat spacetimes are considered, the transverse boundary metric $q_{AB}$ also plays the role of source \cite{Campiglia:2014yka , Campiglia:2015yka , Compere:2018ylh , Campiglia:2020qvc,Flanagan:2019vbl}. Freezing the sources eliminates the flux and the charges become integrable. The condition of freezing the sources, $\delta C_{AB} = 0$ ($\delta q_{AB} = 0$), yields a similar symmetry breaking 
\begin{equation}
\begin{aligned}
\begin{tikzpicture}[baseline=(current  bounding  box.center)]
\coordinate (A) at (0,0);
\coordinate (B) at (3,0);
\draw (A) node[left]{\normalsize $\text{(Generalized) BMS}_{d+1}$};
\draw[-latex] (0,0) -- (B);
\draw (B) node[right]{\normalsize $SO(d,1) \loplus \mathbb{R}^4$};
\draw ($(A)!0.5!(B)$) node[above]{\scriptsize freezing sources};
\end{tikzpicture}
\label{Symmetry breaking flat space}
\end{aligned}
\end{equation}
where $SO(d,1) \loplus \mathbb{R}^4$ is the Poincaré algebra. In other words, the infinite-dimensional (generalized) $\text{BMS}_{d+1}$ algebra reduces to the finite-dimensional symmetry group of the Minkowski vacuum when turning off the sources. The symmetry breaking \eqref{Symmetry breaking flat space} is the flat limit of the symmetry breaking \eqref{symmetry breaking AdS}. These two diagrams commute, namely 

\begin{equation}
\begin{aligned}
\begin{tikzpicture}[baseline=(current  bounding  box.center)]
\node[] at (-5,0) (A)  {\normalsize $\Lambda\text{-BMS}_{d+1}$};
\node[] at (0,1) (B) {\normalsize $SO(d,2)$};
\node[] at (0,-1) (C) {\normalsize $\text{(Generalized) BMS}_{d+1}$};
\node[] at (5,0) (D) {\normalsize $SO(d,1) \loplus \mathbb{R}^4$.};

\draw[-latex] (A) |- (B);
\draw[-latex,densely dashed] (A) |- (C);
\draw[latex-,densely dashed] (D) |- (B);
\draw[latex-] (D) |- (C);

\draw[] let \p1 = (A), \p2 = (B) in (\x1,\y2) node[above right]{\scriptsize freezing sources};

\draw[] let \p1 = (A), \p2 = (C) in (\x1,\y2) node[below right]{\scriptsize flat limit $\ell\to +\infty$};

\draw[] let \p1 = (D), \p2 = (B) in (\x1,\y2) node[above left]{\scriptsize flat limit $\ell\to +\infty$};

\draw[] let \p1 = (D), \p2 = (C) in (\x1,\y2) node[below left]{\scriptsize freezing sources};

\end{tikzpicture}
\end{aligned}
\vspace{5pt}
\end{equation}

From these considerations, it seems to us that investigating holography in AdS with a dissipative system coupling to an environment would be a excellent starting point to apprehend flat holography in $d>2$. More specifically, it would be interesting to understand how the recent advances in the celestial sphere CFT description \cite{Donnay:2018neh , Ball:2019atb , Nguyen:2020hot , Pate:2019lpp , Donnay:2020guq , Donnay:2020fof , Gonzalez:2020tpi , Himwich:2020rro, Casali:2020vuy , Fotopoulos:2020bqj , Fotopoulos:2019vac , Adamo:2019ipt, Laddha:2020kvp , Lowe:2020qan , Banerjee:2020kaa} are related to the AdS results through a flat limit process.   

Let us notice that while the flat limit is perfectly defined at the level of the symmetries, it is not yet clear how it works at the level of the phase space for arbitrary dimensions. In three dimensions, this limit was worked out in \cite{Barnich:2012aw} (see also \cite{Campoleoni:2018ltl , Ciambelli:2020eba , Ciambelli:2020ftk}). In four dimensions, the flat limit of the phase space was achieved in \cite{Compere:2020lrt}, thanks to a diffeomorphism between Bondi and Starobinsky/Fefferman-Graham gauges \cite{Poole:2018koa , Compere:2019bua} (see also \cite{Ruzziconi:2019pzd , Ruzziconi:2020cjt}). This analysis requires a complete control on the solution space in asymptotically (A)dS spacetime, in a gauge that owns a well-defined flat limit $\ell \to \infty$, \textit{e.g.} the Bondi gauge. Up to our knowledge, such a framework is not yet available for arbitrary dimensions. In particular, it would be interesting to see how the obstructions for conformal compactification in odd-dimensional asymptotically flat spacetimes (with $d> 2$) arise in the flat limit \cite{Hollands:2004ac}.   

Finally, let us emphasize that our analysis has been performed in the gauge-fixing approach \cite{Ruzziconi:2019pzd}, \textit{i.e.} we have fixed the Starobinsky/Fefferman-Graham gauge before imposing the boundary conditions. As suggested in \cite{Grumiller:2016pqb , Grumiller:2017sjh , Ciambelli:2020ftk , Ciambelli:2020eba}, additional symmetries may be uncovered by relaxing the gauge fixing conditions before imposing the falloffs on the metric components. In particular, partial gauge fixings may be more adapted to interpret our results concerning the features of Weyl symmetries in odd dimensions and their interplay with holography \cite{Ciambelli:2019bzz , Ciambelli:2020ftk , Ciambelli:2020eba}.

\section*{Acknowledgements}

We are indebted to Geoffrey Comp\`ere for precious comments on the manuscript and for paving the way towards this project. We thank Frank Ferrari for pointing us some useful references and for a very illuminating conversation. We also thank Francesco Alessio, Glenn Barnich, Luca Ciambelli, Daniel Grumiller, Yannick Herfray, Charles Marteau, Pujian Mao, Marios Petropoulos and C\'eline Zwikel for useful discussions and collaborations on related subjects. AF was supported by the Fonds de la Recherche Scientifique F.R.S.-FNRS (Belgium). RR was partially supported by the FRIA (F.R.S.-FNRS, Belgium) and by the Austrian Science Fund (FWF), project P 32581-N.

\appendix

\section{Review of the solution space}
\label{Review of the solution space}

This appendix aims to review how to derive the solutions of the Einstein field equations in the Starobinsky/Fefferman-Graham gauge for the leading orders we are interested in. The review is mainly based on \cite{deHaro:2000vlm}, up to some changes of conventions, but the solving procedure is slightly different.

\paragraph{Organization of the Einstein equations} In Starobinsky/Fefferman-Graham gauge $g_{\rho\rho} = \eta \frac{\ell^2}{\rho^2}$, $g_{\rho a} = 0$, the coefficients of the Levi-Civita connection can be expressed in terms of the induced metric $\gamma_{ab}(\rho,x^c)$ on constant $\rho$ hypersurfaces as
\begin{equation}
\begin{split}
&\Gamma^\rho_{\rho\rho} = -\frac{1}{\rho}, \quad \Gamma^\rho_{\rho a} = 0,\quad \Gamma^a_{\rho\rho} = 0, \\
&\Gamma^a_{b\rho} = \frac{1}{2}\gamma^{ac}\partial_\rho \gamma_{bc}, \quad \Gamma^\rho_{ab} = -\frac{1}{2}\eta \frac{\rho^2}{\ell^2}\partial_\rho \gamma_{ab}, \quad \Gamma^a_{bc} = \Gamma^a_{bc}[\gamma].
\end{split}
\end{equation}
Note that $\Gamma^{a}_{\rho a} = -\eta \frac{\ell^2}{\rho^2}\gamma^{ab}\Gamma^\rho_{ab}$, a very useful property for practical computations. Denoting by $\mathcal D_a$ the Levi-Civita connection associated with $\gamma_{ab}(\rho,x^c)$, the components of the Ricci tensor read as
\begin{equation}
\begin{split}
R_{ab} &= R_{ab}[\gamma] + (\partial_\rho+\Gamma^\rho_{\rho\rho})\Gamma^\rho_{ab} + \Gamma^\rho_{ab} \Gamma^c_{c\rho} - 2 \Gamma^{\rho}_{c(a}\Gamma^c_{b)\rho} \\
&= R_{ab}[\gamma] - \frac{1}{2}\eta \frac{\rho^2}{\ell^2} \left[ \Big(\partial_\rho  + \frac{1}{\rho}\Big)\partial_\rho \gamma_{ab} - \partial_\rho \gamma_{c(a}\gamma^{cd}\partial_\rho \gamma_{b)d} + \frac{1}{2}(\gamma^{cd}\partial_\rho \gamma_{cd})\partial_\rho \gamma_{ab} \right], \\
R_{\rho a} &= -2 \mathcal D_{[a} \Gamma^b_{b]\rho} = \frac{1}{2}\left[ \mathcal D^b (\partial_\rho \gamma_{ab}) - \mathcal D_{a} (\gamma^{bc}\partial_\rho \gamma_{bc}) \right], \\
R_{\rho\rho} &= -(\partial_\rho - \Gamma^\rho_{\rho\rho})\Gamma^a_{a\rho} - \Gamma^{a}_{b\rho}\Gamma^b_{a\rho} = -\frac{1}{2}\left( \partial_\rho + \frac{1}{\rho} \right) (\gamma^{ab}\partial_\rho \gamma_{ab}) - \frac{1}{4}\gamma^{ac}\gamma^{bd}\partial_\rho \gamma_{ab}\partial_\rho \gamma_{cd}  .
\end{split}
\end{equation}
The Ricci curvature is
\begin{equation}
R = \gamma^{ab} R_{ab} [\gamma] - \eta\frac{\rho^2}{\ell^2} \left[ \gamma^{ab}\left(\partial_\rho  + \frac{1}{\rho}\right)\partial_\rho\gamma_{ab} - \frac{3}{4}\gamma^{ac}\gamma^{bd}\partial_\rho\gamma_{ab}\partial_\rho\gamma_{cd} + \frac{1}{4}(\gamma^{ab}\partial_\rho\gamma_{ab})^2\right].
\end{equation}
Due to the particular form of the bulk metric, Einstein's equations read as follows: \begin{align}
R_{\rho\rho} &= -\frac{d}{\rho^2}  , \label{Einstein rhorho} \\
R_{\rho a} &= 0 , \label{Einstein rhoa} \\
R_{ab} &= -\eta\frac{d}{\ell^2}\gamma_{ab} \label{Einstein ab}
\end{align}
where the Ricci curvature is set to its on-shell value
\begin{equation}
R = -\eta \frac{d(d+1)}{\ell^2}  . \label{Einstein supp}
\end{equation}
These four equations can be solved order by order in $\rho$ if $\gamma_{ab}(\rho,x^c)$ is expressed as 
\begin{equation}
\gamma_{ab} = \frac{1}{\rho^2} \left( g^{(0)}_{ab} + \rho g^{(1)}_{ab} + \rho^2 g^{(2)}_{ab} + \rho^3 g^{(3)}_{ab} + \dots + \rho^d g^{(d)}_{ab} + \rho^d \ln \rho^2 \tilde g^{[d]}_{ab} + \cdots \right)
\end{equation}
where the logarithmic term is introduced for even $d$ only. We can already eliminate $g_{ab}^{(1)}$ since the leading order $\mathcal O(1/\rho)$ of \eqref{Einstein ab} imposes that $(d-1)g^{(1)}_{ab} = 0$, hence this field is zero for any interesting case $d\geq 2$. 
 
The purely radial constraint \eqref{Einstein rhorho} at order $\mathcal O(\rho^{k-2})$ fixes the trace of $g_{ab}^{(k)}$ with respect to $g^{(0)}_{ab}$ for $k\geq 3$. The $\mathcal O(\rho^0)$ of the equation is tautologic, hence the first non-trivial trace, $g_{(0)}^{ab}g^{(2)}_{ab}$, is not actually fixed by \eqref{Einstein rhorho} and requires a particular treatment. In fact, using $R[\gamma] = \rho^2 R^{(0)}+\mathcal O(\rho^3)$, that quantity can be extracted easily from \eqref{Einstein supp} at leading order
\begin{equation}
\boxed{
g^{ab}_{(0)} g^{(2)}_{ab} = -\frac{\eta\,\ell^2 }{2(d-1)}R^{(0)}. \label{Tr g2}
}
\end{equation}
For any dimension $d$, the leading logarithmic term in \eqref{Einstein rhorho} imposes that $\tilde g^{[d]}_{ab}$ is trace-free. 

Once the traces of the various coefficients have been fixed, the momentum constraint \eqref{Einstein rhoa} at order $\mathcal O(\rho^{k-1})$ fixes the covariant divergence of $g^{(k)}_{ab}$ with respect to $g^{(0)}_{ab}$ for $k\geq 2$. In particular, $D^b g^{(k)}_{ab} = 0$ is identically zero when $k$ is odd. For any dimension $d$, the leading logarithmic term in \eqref{Einstein rhoa} imposes that $\tilde g^{[d]}_{ab}$ is divergence-free. 

Once the traces and divergences of the various coefficients have been fixed by \eqref{Einstein rhorho} and \eqref{Einstein rhoa}, the transverse components \eqref{Einstein ab} are algebraically solved to get the explicit expressions of the aforementioned coefficients in terms of lower order coefficients. Starting from $g^{(1)}_{ab}=0$, one can show inductively that $(d-k)g^{(k)}_{ab} = 0$ for $k$ odd at order $\mathcal O(\rho^{k-2})$, hence only even powers of $\rho$ appear in \eqref{preliminary FG} up to $\mathcal O(\rho^{2[\frac{d-1}{2}]})$ where $[x]$ indicates the integer part of $x$. For $k$ even, one gets generically $(d-k)g^{(k)}_{ab} = \mathcal C^{(k)}[g^{(0)},...,g^{(k-1)}]$ for $k < d$, ${\mathcal C'}^{(d)}[g^{(0)},...,g^{(d-1)},\tilde g^{[d]}]=0$ for $k=d$ and $(d-k)g^{(k)}_{ab} = {\mathcal C''}^{(k)}[g^{(0)},...,g^{(d)},\tilde g^{[d]},...,g^{(k-1)}]$ for $k>d$. In any dimension $d$, we see that the expression $g_{ab}^{(d)}$ is always left unfixed by the equations of motion, only its trace and divergence can be computed from \eqref{Einstein rhorho} and \eqref{Einstein rhoa} respectively. We also remark that the logarithmic piece $\rho^{(d-2)}\ln\rho^2 \tilde g^{[d]}_{ab}$ is necessary to ensure the consistency of Einstein's equations when $d$ is even. Indeed, if we do not introduce the field $\tilde g^{[d]}_{ab}$, the equation $\tilde{\mathcal C}^{(d)}[g^{(0)},...,g^{(d-1)}]=0$ brings in general an additional constraint on the coefficients $g^{(0)}_{ab},...,g^{(d-1)}_{ab}$ which are supposed to have been fixed at lower order in $\rho$. Rather than that, ${\mathcal C'}^{(d)}[g^{(0)},...,g^{(d-1)},\tilde g^{[d]}]=0$ gives in these cases the explicit solution for $\tilde g^{[d]}_{ab}$ in terms of lower order coefficients. To summarize,
\begin{equation}
\boxed{
\tilde g^{[2k+1]}_{ab} = 0 , \quad \tilde g^{[2k]}_{ab} \neq 0.
}
\end{equation}
A notable exception occurs for $d=2$ as we will see below in more details, simply because ${\mathcal C'}^{(2)}[g^{(0)}]=0$ on-shell without the help of any additional field. This closes our description of the procedure to get the solution space in Starobinsky/Fefferman-Graham gauge.

\paragraph{Trace of metric coefficients} Apart of the trace of $g^{(2)}_{ab}$ set as \eqref{Tr g2}, we extract the various traces from the equation \eqref{Einstein rhorho}, starting at $\mathcal O(\rho)$ order:
\begin{align}
g^{ab}_{(0)} g^{(3)}_{ab} &= 0  , \label{Tr g3} \\
g^{ab}_{(0)} g^{(4)}_{ab} &= \frac{1}{4} g^{ab}_{(2)} g^{(2)}_{ab}   , \label{Tr g4} \\
g^{ab}_{(0)} g^{(5)}_{ab} &= 0   , \label{Tr g5} \\
g^{ab}_{(0)} g^{(6)}_{ab} &= \frac{2}{3} g_{(2)}^{ab}g^{(4)}_{ab} - \frac{1}{6}(g_{(0)}^{ab}g^{(2)}_{ab})^3  ,\  \dots \label{Tr g6}
\end{align}
These equations help to compute the trace of the boundary stress-tensor in \eqref{trace even}--\eqref{trace 4d}.

\paragraph{Divergence of metric coefficients} Developing the momentum constraint \eqref{Einstein rhoa} leads to the generic form $D^{b} g^{(k)}_{ab} = -D^{b} X^{[k]}_{ab} + V_a^{[k]}$. This equation is crucial because it allows to check explicitly that the boundary stress-tensor $T_{ab}^{[d]}$ is indeed divergence-free. For any odd $k$, we simply have $X_{ab}^{[k]} = 0$ and $V_a^{[k]} = 0$. For even $k$, we can show that
\begin{equation}
\begin{split}
X^{[2]}_{ab} &= -(g^{cd}_{(0)} g_{cd}^{(2)})g_{ab}^{(0)}, \quad  V_a^{[2]} = 0, \\
X^{[4]}_{ab} &= -\frac{1}{8}g_{ab}^{(0)} \left[ (g_{(0)}^{cd} g^{(2)}_{cd})^2 - g_{(2)}^{cd} g^{(2)}_{cd}  \right] - \frac{1}{2} g_{(2)}^{ac} g_{(0)}^{cd} g^{(2)}_{bd} + \frac{1}{4} (g_{(0)}^{cd} g^{(2)}_{cd}) g^{(2)}_{ab} ,  \quad V_a^{[4]} = 0,  \ \dots 
\end{split}
\end{equation}
These results show that the combination $g_{ab}^{(d)}+X^{[d]}_{ab}$ is sufficient to get the conservation of the holographic stress-tensor for the dimensions $d\leq 4$, see equation \eqref{divergence free}. One can notice the presence of a non-vanishing $V_a$ covector for $d=6$ and generically beyond. It makes the construction of candidates for $T_{ab}^{[d]}$ from the equations of motion slightly more involved. Moreover, one can also check that $X_{ab}^{[6]}$ contains an anti-symmetric part, which was not the case for lower dimensions. However it was shown in \cite{deHaro:2000vlm}, for the case of $d=6$, that one can perform a field redefinition from $g_{ab}^{(6)}$ in order to get a covariantly conserved symmetrical tensor. 

\paragraph{Explicit expressions of metric coefficients} Now that we have the trace and the divergence of the metric coefficients we are interested in, the last equation \eqref{Einstein ab} allows to derive their explicit form in terms of lower order coefficients. At leading order $\mathcal O(\rho^0)$, we find
\begin{equation}
\boxed{
g_{ab}^{(2)} = -\frac{\eta \ell^2}{d-2} \left( R_{ab}^{(0)} - \frac{R^{(0)}}{2(d-1)}g_{ab}^{(0)}\right), \ \forall d \geq 3. \label{g2}
}
\end{equation}
For $d=2$, the $\mathcal O(\rho^0)$ contribution of \eqref{Einstein ab} does not constrain $g_{ab}^{(2)}$ but fixes the logarithmic term $\tilde g_{ab}^{[2]}$ in the expansion, and states nothing but $\tilde g^{[2]}_{ab} = \frac{1}{2} \eta \ell^2 (R_{ab}^{(0)} - \frac{1}{2}R^{(0)}g_{ab}^{(0)})$, which leads to
\begin{equation}
\boxed{
\tilde g_{ab}^{[2]} = 0,
}
\end{equation}
since the Einstein tensor vanishes identically in two dimensions. At next order $\mathcal O(\rho)$, we find
\begin{equation}
\boxed{
(d-3)g^{(3)}_{ab} = 0,
}
\end{equation}
which means that $g^{(3)}_{ab}$ is identically zero except for $d=3$ where it is unconstrained. Going further at order $\mathcal O(\rho^2)$, we find for $d\neq 4$
\begin{equation}
(d-4)g^{(4)}_{ab} + g_{ac}^{(2)} g_{(0)}^{cd} g_{bd}^{(2)} - \frac{1}{4}g_{(2)}^{cd}g^{(2)}_{cd}g_{ab}^{(0)} + \frac{1}{2}\eta\ell^2 R^{(2)}_{ab}[g^{(0)}] = 0 \label{g4 temp}
\end{equation}
where we noted that $R_{ab}[\gamma] = R^{(0)}_{ab}[g^{(0)}] + \rho^2 R_{ab}^{(2)}[g^{(0)}] + \mathcal O(\rho^3)$. In order to extract some information from this equation, we need to compute the first subleading piece of the Ricci tensor. First of all, we can remark that ${\Gamma^a}_{bc}[\gamma] = {\Gamma^a}_{bc}[g^{(0)}] + \rho^2 \Gamma^a_{(2)bc}[g^{(0)}] + \mathcal O(\rho^3)$, where
\begin{equation}
\begin{split}
\Gamma^a_{(2)bc}[g^{(0)}] &= \frac{1}{2}g_{(0)}^{ad} (\partial_b g_{dc}^{(2)} + \partial_c g_{bd}^{(2)} - \partial_d g^{(2)}_{bc} ) - \frac{1}{2}g_{(2)}^{ad} (\partial_b g_{dc}^{(0)} + \partial_c g_{bd}^{(0)} - \partial_d g^{(0)}_{bc} ) \\
&= \frac{1}{2}g_{(0)}^{ad} (D_b g_{cd}^{(2)} + D_c g_{bc}^{(2)} - D_d g^{(2)}_{bc})
\end{split}
\end{equation}
is a covariant $(1,2)$-tensor with respect to the boundary geometry and only depends on $g^{(0)}_{ab}$ and its curvature because of \eqref{g2}. As a consequence,
\begin{align}
&R_{ab}^{(2)}[g^{(0)}] = D_c \Gamma^c_{(2)ab}[g^{(0)}] - D_b \Gamma^c_{(2)ac}[g^{(0)}] \nonumber \\
&\quad = D^c D_{(a} g^{(2)}_{b)c} - \frac{1}{2}D^2 g_{ab}^{(2)} - \frac{1}{2}D_aD_b (g^{cd}_{(0)}g^{(2)}_{cd}) \\
&\quad = -\frac{\eta\,\ell^2}{d-2}\left[R_{ca}^{(0)}R^{c}_{(0)b} - R_{acbd}^{(0)}R^{cd}_{(0)} + \frac{(d-2)}{4(d-1)}D_aD_bR^{(0)} - \frac{1}{2} D^c D_c R_{ab}^{(0)} + \frac{D^c D_c R^{(0)}}{4(d-1)} g^{(0)}_{ab} \right].\nonumber 
\end{align}
We can deduce that
\begin{equation}
R[\gamma] = \gamma^{ab}R_{ab}[\gamma] = \rho^2 R^{(0)} + \rho^4 \frac{\eta\,\ell^2}{(d-2)}\left[ R^{ab}_{(0)}R_{ab}^{(0)} - \frac{1}{2(d-1)}(R^{(0)})^2\right] + \mathcal O(\rho^5).
\end{equation}
Finally computing
\begin{equation}
\begin{split}
g_{(2)a}^{c} g_{bd}^{(2)} &= \frac{\ell^4}{(d-2)^2}\left[ R_{(0)a}^{c} R_{bd}^{(0)} - \frac{R^{(0)}}{d-1}R_{ab}^{(0)} + \frac{(R^{(0)})^2}{4(d-1)^2}g^{(0)}_{ab} \right], \\
g_{(2)}^{cd}g^{(2)}_{cd} &= \frac{\ell^4}{(d-2)^2}\left[ R^{cd}_{(0)} R_{cd}^{(0)} + \frac{(4-3d)}{4(d-1)^2}(R^{(0)})^2\right],
\end{split} \label{g2square}
\end{equation}
we can solve \eqref{g4 temp} for $g^{(4)}_{ab}$ if $d\neq 4$ to get
\begin{equation}
\boxed{
\begin{aligned}
g^{(4)}_{ab} = \frac{\ell^4}{(d-4)}  & \left[ \frac{1}{8(d-1)}D_a D_b R^{(0)} - \frac{1}{4(d-2)}D_c D^c R_{ab}^{(0)} + \frac{D_c D^c R^{(0)}}{8(d-1)(d-2)}g_{ab}^{(0)} \right.\\
& \quad - \frac{1}{2(d-2)}R_{acbd}^{(0)}R^{cd}_{(0)} + \frac{(d-4)}{2(d-2)^2}R_a^{(0)c} R^{(0)}_{cb} + \frac{R^{(0)}}{(d-1)(d-2)^2}R_{ab}^{(0)} \\
& \quad  \left. +\, \frac{1}{4(d-2)^2}R^{cd}_{(0)} R_{cd}^{(0)} g^{(0)}_{ab} - \frac{3d}{16(d-1)^2(d-2)^2}(R^{(0)})^2 g^{(0)}_{ab} \right] . 
\end{aligned} \label{g4}
}
\end{equation}
When $d=4$, \eqref{g4 temp} fixes the logarithmic term $\tilde g^{[4]}_{ab}$ in the expansion. After some algebra we obtain
\begin{equation}
\begin{split}
\tilde g^{[4]}_{ab} &= \frac{1}{2} g_{ac}^{(2)}g^{cd}_{(0)}g^{(2)}_{bd} - \frac{1}{8} g_{(2)}^{cd} g^{(2)}_{cd} g^{(0)}_{ab} + \frac{\eta\ell^2}{4}R^{(2)}_{ab}  \\
&= \frac{1}{2} g_{ac}^{(2)}g^{cd}_{(0)}g^{(2)}_{bd} - \frac{1}{8} g_{(2)}^{cd} g^{(2)}_{cd} g^{(0)}_{ab} + \frac{\eta\ell^2}{8}\left[ 2 D^c D_{(a} g^{(2)}_{b)c} - D^2 g_{ab}^{(2)} - D_aD_b (g^{cd}_{(0)}g^{(2)}_{cd}) \right],
\end{split}
\end{equation}
or explicitly in terms of $g^{(0)}_{ab}$
\begin{equation}
\boxed{
\begin{aligned}
\tilde g^{[4]}_{ab} &= \frac{\ell^4}{8}R_{acbd}^{(0)}R^{cd}_{(0)} - \frac{\ell^4}{48}D_aD_b R^{(0)} + \frac{\ell^4}{16}D_c D^c R_{ab}^{(0)} - \frac{\ell^4}{24}R^{(0)}R^{(0)}_{ab}  \\
&\quad\ + \frac{\ell^4}{96}\left[ (R^{(0)})^2 - D_c D^c R^{(0)} - 3 R^{cd}_{(0)} R^{(0)}_{cd}\right]g^{(0)}_{ab}. 
\end{aligned}
} \label{tilde 4}
\end{equation}
The algorithm can be pursued as follows. We check that $(d-5)g^{(5)}_{ab}$ is linear in $g^{(3)}_{ab}$, so we see that $g^{(5)}_{ab}=0$ except for $d=5$ as expected. We have $(d-6)g^{(6)}_{ab} = \mathcal C^{(6)}[g^{(0)}]$ for $d<6$, $(d-6)g^{(6)}_{ab} = \mathcal C^{(6)}[g^{(0)},\tilde g^{(6)}_{ab}]$ for $d>6$, and the same equation fixes $\tilde g^{(6)}_{ab}$ in terms of $g^{(0)}_{ab}$ when $g^{(2)}_{ab}$ and $g^{(4)}_{ab}$ hit their on-shell values. We will not present the results in details here, since we do not need more than what we have derived so far.

\section{Algebra of residual gauge diffeomorphisms}
\label{app:Algebra of residual gauge diffeomorphisms}

In this appendix, we prove the result \eqref{eq:VectorAlgebra}. To simplify the notations, we write $\xi_1 \equiv \xi (\sigma_1, \bar\xi_1)$ and $\xi_2 \equiv \xi (\sigma_2, \bar\xi_2)$. Since these vector fields are residual gauge diffeomorphisms of the Starobinsky/ Fefferman-Graham gauge \eqref{FG gauge}, they satisfy \eqref{eq:CstVector2}. As a result, the computation of $[\xi_1,\xi_2]^\rho_\star$ is straightforward and gives
\begin{equation}
\frac{1}{\rho}[\xi_1,\xi_2]^\rho_\star = \left( \xi^a_1 \partial_a\sigma_2 - \xi^a_2 \partial_a\sigma_1\right) - \delta_{\xi_1}\sigma_2 + \delta_{\xi_2}\sigma_{1}.
\end{equation}
Taking a derivative with respect to $\rho$, and again using \eqref{eq:CstVector2}, we get
\begin{equation}
\partial_\rho \left( \frac{1}{\rho}[\xi_1,\xi_2]^\rho_\star \right) = \partial_\rho \xi^a_1 \partial_a \sigma_2 -\partial_\rho \xi_2^a\partial_a\sigma_{1} = 0,
\end{equation}
which shows that $[\xi_1,\xi_2]^\rho_\star = \rho \hat \sigma$, and
\begin{equation}
\hat \sigma = \frac{1}{\rho}[\xi_1,\xi_2]^\rho_\star \Big|_{\rho=0} =  \bar\xi_1^a\partial_a \sigma_2 - \bar\xi_2 \partial_a \sigma_1 - \delta_{\xi_1} \sigma_2 + \delta_{\xi_2}\sigma_1.
\end{equation}
Let us now consider the transverse components. By evaluating the commutator at leading order, we derive that
\begin{equation}
\hat{\bar \xi}^a = \lim_{\rho\to 0} [\xi_1,\xi_2]^a_\star = [\bar\xi_1,\bar\xi_2]^a - \delta_{\xi_1} \bar\xi_2^a + \delta_{\xi_2} \bar\xi_1^a.
\end{equation}
Recalling that $\delta_\xi \gamma^{ab} = \mathcal L_\xi \gamma^{ab} = \rho\sigma_\xi \partial_\rho \gamma^{ab} + \xi^c \partial_c \gamma^{ab} - 2 \gamma^{c(a}\partial_c \xi^{b)}$ and explicitly using \eqref{eq:CstVector2} to express $\partial_\rho \xi^a_1$ and $\partial_\rho \xi^b_2$ in terms of $\sigma_1$ and $\sigma_2$, respectively, a direct computation yields
\begin{equation}
\partial_\rho \left( [\xi_1,\xi_2]^a_\star \right) = -\eta\ell^2\frac{1}{\rho}\gamma^{ab} \partial_b \hat \sigma.
\end{equation}
We have just proven that residual gauge diffeomorphisms $\xi_1$ and $\xi_2$ of the Starobinsky/Fefferman-Graham gauge satisfy 
\begin{equation}
[\xi(\sigma_1,\bar\xi_1^a),\xi(\sigma_2,\bar\xi_2^a)]_\star =  \xi (\hat \sigma,\hat{\bar\xi}^a) 
\end{equation}
where
\begin{equation}
\begin{split}
\hat \sigma &= \bar\xi_1^a\partial_a \sigma_2  - \delta_{\xi_1} \sigma_2 - (1 \leftrightarrow 2 ), \\
\hat{\bar\xi}^a &=  \bar\xi_1^b \partial_b \bar\xi_2^a - \delta_{\xi_1} \bar\xi_2^a - (1 \leftrightarrow 2).
\end{split} 
\end{equation}

\section{Variations of the solution space}
\label{Variations of the solution space}

In this appendix, we provide some intermediate steps to obtain the variations of the solution space \eqref{eq:action solution space} and \eqref{eq:action solution space 2} under residual gauge diffeomorphisms. Up to conventions, this derivation is the same as the one provided in \cite{Imbimbo:1999bj , deHaro:2000vlm}. 

Writing \eqref{AKV 2} as $\xi^a(\rho,x^b) = \bar\xi^a(x^b) + \sum_{k=1}^{+\infty} \xi^a_{(k)}(\rho,x^b) \rho^k$, we have
\begin{equation}
\xi^a_{(1)} = 0, \quad \xi^a_{(2)} = -\frac{1}{2}\eta\ell^2 g^{ab}_{(0)}\partial_b \sigma, \quad \xi^a_{(3)} = 0, \quad \xi^a_{(4)} = \frac{1}{4}\eta\ell^2 g^{ab}_{(2)}\partial_b \sigma.
\end{equation}
Under the full residual gauge diffeomorphisms $\xi = \xi^\rho\partial_\rho + \xi^a \partial_a$, the metric varies as
\begin{equation}
\delta_\xi \gamma_{ab} (\rho,x^c) = \left( \sigma\rho\partial_\rho + \mathcal L_{\xi^c} \right) \gamma_{ab}(\rho,x^c).
 \end{equation}
Expanding this in power series of $\rho$ yields
\begin{align}
\delta_\xi g^{(0)}_{ab} &= \mathcal L_{\bar\xi} g^{(0)}_{ab} - 2\sigma g^{(0)}_{ab}, \\
\delta_\xi g^{(2)}_{ab} &= \mathcal L_{\bar\xi} g^{(2)}_{ab} +2  D_{(a} \xi^{(2)}_{b)}, \\
\delta_\xi g^{(3)}_{ab} &= \mathcal L_{\bar\xi} g^{(3)}_{ab} + \sigma g^{(3)}_{ab}, \\
\delta_\xi g^{(4)}_{ab} &= \mathcal L_{\bar\xi} g^{(4)}_{ab} + 2 \sigma (g^{(4)}_{ab}+\tilde g^{[4]}_{ab})+ 2 D_{(a} \xi_{b)}^{(4)} + \xi_{(2)}^c D_c g^{(2)}_{ab}  + 2 g^{(2)}_{c(a} D_{b)}\xi^c_{(2)}.
\end{align} From these variations, one can easily extract \eqref{eq:action solution space} and \eqref{eq:action solution space 2} by considering the appropriate orders. In particular, the above variations reproduce those obtained under PBH transformations when setting $\bar{\xi}^a = 0$ \cite{Imbimbo:1999bj}.

\section{Holographic renormalization}
\label{app:Holographic renormalization}

In this appendix, we review and detail some intermediate computations for the holographic renormalization procedure outlined in Section \ref{sec:Holographic renormalization} \cite{deHaro:2000vlm , Bianchi:2001kw}. The standard variational principle for Einstein gravity on a $(d+1)$-dimensional manifold $\mathscr M$ with boundary $\mathscr I$ is given by 
\begin{equation}
S = \int_{\mathscr{M}}  \bm L_{EH} + \int_{\mathscr{I}}  \bm L_{GHY}
\end{equation}
where $\bm L_{EH}$ is the Einstein-Hilbert Lagrangian \eqref{LEH} and $\bm L_{GHY}$ the usual Gibbons-Hawking-York boundary term \eqref{LGHY}. In order to evaluate the on-shell action and deal with the radial divergences that will appear when approaching the boundary $\mathscr I$, we introduce a regulator $\epsilon>0$ and define the regularized action
\begin{equation}
S_{reg}^\epsilon = \int_{\rho \ge \epsilon} \bm L_{EH} + \int_{\rho = \epsilon}  \bm L_{GHY} 
\end{equation}
When the equations of motion hold, we have $R = 2\Lambda \frac{(d+1)}{(d-1)} = -\eta \frac{d(d+1)}{\ell^2}$, and $\sqrt{-g} = \frac{\ell}{\rho}\sqrt{|\gamma|}$. The second fundamental form of $\mathscr I$ is defined as $K_{ab} = \nabla_{(a} N_{b)}$ where $N_\mu = -\eta\sqrt{|g_{\rho\rho}|}\delta^\rho_\mu$ are the covariant components of the outward normal vector $N$. Hence the extrinsic curvature reads as $K = \gamma^{ab}\nabla_a N_b = -\gamma^{ab} \Gamma^\rho_{ab}N_\rho = \frac{1}{2}g^{\rho\rho}\gamma^{ab}\partial_\rho\gamma_{ab}N_\rho$, allowing us to derive the expression of $S_{reg}^\epsilon$ in terms of the boundary volume form only:
\begin{equation}
S_{reg}^\epsilon = \frac{\eta}{16\pi G\ell} \int \D^d x \left[ \int_{\epsilon}^\infty d\rho \left( -\frac{2d}{\rho} \sqrt{|\gamma|} \right) - \Big[ 2\rho\partial_\rho \sqrt{|\gamma|} \Big]\Big|_{\rho=\epsilon} \right]. \label{Sreg}
\end{equation}
Note that the upper bound $\infty$ of the integral should be intended as some cut-off in the bulk within the validity range of the Starobinsky/Fefferman-Graham coordinates, and all such contributions will be ignored, as usual, in the formulation of the variational principle. Now the evaluation of the radial divergences amounts to plug the polyhomogeneous expansion of $\sqrt{|\gamma|} = \rho^{-d}\sqrt{|g^{(0)}|} \sqrt{\det(\delta^c_b+h^c_b)}$, $h^c_b = \mathcal O(\rho^2)$ into \eqref{Sreg}. We get $S_{reg}^{\epsilon} = S_{reg}^{\epsilon,div}[g^{(0)}] + \mathcal O(\varepsilon^0)$ with
\begin{equation}
S_{reg}^{\epsilon,div}[g^{(0)}] = \frac{\eta}{16\pi G\ell} \int \D^d x \frac{\sqrt{|g^{(0)}|} }{\epsilon^d} \left( a_{(0)} + \epsilon^2 a_{(2)} + \epsilon^4 a_{(4)} + \dots + \epsilon^{d-2} a_{(d-2)} - \ln \epsilon^2\,  \tilde a_{[d]} \right). \label{Radial div}
\end{equation}
The coefficients $a_{(k)}$, $k\neq d$, are non-trivial only for even $k$. The logarithmic divergence is due to the integration in the first term of \eqref{Sreg} when one considers the finite part of $\sqrt{|\gamma|}$ times $1/\rho$. Since this part is shown to be precisely the conformal anomaly in $d$ dimensions (up to some numerical factor), the coefficient $\tilde a_{[d]}$ appears thus only for even $d$. The explicit expressions are
\begin{equation}
\begin{split}
a_{(0)} &= 2(d-1) \ , \\
a_{(2)} &= \frac{(d-4)(d-1)}{(d-2)}(g^{ab}_{(0)}g^{(2)}_{ab})  , \\
a_{(4)} &= \frac{16-9d+d^2}{4(d-4)}\left[ (g_{(0)}^{ab} g^{(2)}_{ab})^2 - g_{(2)}^{ab} g^{(2)}_{ab} \right]  , \ \dots
\end{split}
\end{equation}
and 
\begin{equation}
\tilde a_{[2]} = -(g^{ab}_{(0)}g^{(2)}_{ab}) , \quad \tilde a_{[4]} = -\frac{1}{2}\left[ (g_{(0)}^{ab} g^{(2)}_{ab})^2 - g_{(2)}^{ab} g^{(2)}_{ab} \right]  , \ \dots
\end{equation}
One should make good use of \eqref{Tr g4} to get the final form of $a_{(4)}$ and $\tilde a_{[4]}$. As a consequence, the regulated variational principle must be supplied by a counter-term action such that
\begin{equation}
S_{ren}^\epsilon \equiv  S_{reg}^\epsilon + S_{ct}^\epsilon   , \quad \  S_{ct}^\epsilon= \int_{\rho=\epsilon} \D^dx \,  \bm L_{ct}[\gamma;\epsilon]  
\end{equation}
is finite and the limit $\epsilon\to 0$ can be safely taken, which yields the renormalized action $S_{ren} = \lim_{\epsilon\to 0}S_{ren}^\epsilon = \mathcal O(\epsilon^0)$. Although $\bm L_{ct}[\gamma;\epsilon]$ can be inferred directly from \eqref{Radial div}, the resulting expression will not reveal the fact that the counter-term Lagrangian is in fact a covariant object with respect to the induced metric $\gamma_{ab}(\epsilon,x^c)$. But one can guess the form of $\bm L_{ct}$ in terms of the volume form $\sqrt{|\gamma|}$ and a power series in the Ricci tensor $R_{ab}[\gamma]$. We give here the first pieces of the development:
\begin{equation}
\begin{split}
\bm L_{ct}[\gamma;\epsilon] = \frac{1}{16\pi G}\frac{\eta}{\ell} \Big[ & -2(d-1)\sqrt{|\gamma|} - \frac{\eta \ell^2}{(d-2)}R[\gamma]\sqrt{|\gamma|} \\
& - \frac{\ell^4}{(d-4)(d-2)^2}\sqrt{|\gamma|}\left(R^{ab}[\gamma]R_{ab}[\gamma] - \frac{d}{4(d-1)}R[\gamma]^2\right) \\
& + \sqrt{|\gamma|}\, \tilde a_{[d]} \ln \epsilon^2 + \mathcal O(R[\gamma]^3) \Big]  (\D^d x)  .
\end{split} \label{Lct in terms of gamma}
\end{equation}
One can check by a straighforward computation, involving \eqref{Tr g2}, \eqref{g2square} and 
\begin{equation}
(g_{(0)}^{ab} g^{(2)}_{ab})^2 - g_{(2)}^{ab} g^{(2)}_{ab} = -\frac{\ell^4}{(d-2)^2}\left[R^{ab}_{(0)}R^{(0)}_{ab} - \frac{d}{4(d-1)}(R^{(0)})^2\right]  ,
\end{equation}
that \eqref{Lct in terms of gamma} encompasses the boundary-covariant counter-terms needed to subtract the radial divergences in \eqref{Radial div} up to $d=6$, while the neglected $\mathcal O(R[\gamma]^3)$ are intended for renormalizing in higher dimensions. Obviously, the number of terms needed for each $d$ depends on $d$, and one does not have to worry about the fact that the second and third terms are singular for $d=2$ and $d=4$ respectively, because they are not supposed to be introduced for these particular dimensions. For $d=2k$, the renormalization requires the participation of the first $k$ counter-terms only, together with the logarithmic divergence. For $d=2k+1$, $\tilde a_{[d]}$ is set to zero by the Einstein equations and only the first $k+1$ counter-terms participate to the renormalization procedure. 

Now we can define the holographic stress-tensor as \cite{Balasubramanian:1999re,deHaro:2000vlm}
\begin{equation}
T_{ab}^{[d]} \equiv -\frac{2}{\sqrt{|g^{(0)}|}} \frac{\partial S_{ren}}{\partial g^{ab}_{(0)}} = \lim_{\epsilon \to 0} \left( \frac{1}{\rho^{d-2}}\frac{-2}{\sqrt{|\gamma|}} \frac{\partial S^\epsilon_{ren}}{\partial \gamma^{ab}}\Big|_{\rho=\epsilon} \right) \equiv \lim_{\epsilon \to 0} \left( \frac{1}{\rho^{d-2}} \texttt T_{ab}^{[d]}[\gamma]\Big|_{\rho=\epsilon}\right)
\end{equation}
where $\texttt T_{ab}^{[d]}[\gamma]$ designates the stress-energy tensor of the theory living on the $\rho=\epsilon$ hypersurface described by $S_{ren}^\epsilon$, and computed with respect to the induced metric $\gamma_{ab}(\epsilon,x^c)$. This comes up with three pieces $\texttt T_{ab}^{[d]}[\gamma] = \texttt T_{ab}^{reg,[d]}[\gamma] + \texttt T_{ab}^{ct,[d]}[\gamma] - \ln \epsilon^2 \, \texttt T_{ab}^{log,[d]}[\gamma]$ where
\begin{equation}
\texttt T_{ab}^{reg,[d]}[\gamma] = -\frac{2}{\sqrt{|\gamma|}} \frac{\partial S^\epsilon_{reg}}{\partial \gamma^{ab}}\Big|_{\rho=\epsilon} = \frac{\eta}{8\pi G} ( K_{ab} - K\,\gamma_{ab} ) \Big|_{\rho=\epsilon}
\end{equation}
is the contribution from the regulated Einstein-Hilbert action,
\begin{equation}
\begin{split}
\texttt T_{ab}^{ct,[d]}[\gamma] &=  -\frac{2}{\sqrt{|\gamma|}} \frac{\partial S^\epsilon_{ct}}{\partial \gamma^{ab}}\Big|_{\rho=\epsilon} \\
&= \frac{1}{8\pi G}\frac{\eta}{\ell} \left[ -(d-1)\gamma_{ab} + \frac{\eta\,\ell^2}{(d-2)}\left(R_{ab}[\gamma] - \frac{1}{2}R[\gamma]\gamma_{ab}\right) + \mathcal O(R[\gamma]^2) \right] 
\end{split}
\end{equation}
is the piece from the counter-term action, and finally
\begin{equation}
\texttt T_{ab}^{log,[d]}[\gamma] = -\frac{2}{\sqrt{|\gamma|}} \frac{\delta (\sqrt{|\gamma|}\,\tilde a_{[d]})}{\delta \gamma^{ab}}\Big|_{\rho=\epsilon} 
\end{equation}
is the stress-tensor associated with the action whose integral kernel is the conformal anomaly (up to a numerical factor). The terms we are interested in run up to the $\mathcal O(\epsilon^2)$ order. We can develop order by order, checking at each step that all the divergences cancel out, and finally picking the leading order. For $d=2$, one gets immediately
\begin{equation}
\texttt T^{[2]}_{ab}[\gamma] = \frac{\eta}{8\pi G \ell} \left[ g^{(2)}_{ab} - (g^{cd}_{(0)} g^{(2)}_{cd})g^{(0)}_{ab} \right] + \mathcal O(\epsilon^2)
\end{equation}
which verifies \eqref{holographic stress-energy tensor} and \eqref{X2}. For $d=3$, the only surviving term is
\begin{equation}
\texttt T^{[3]}_{ab}[\gamma] = \epsilon\frac{3\eta}{16\pi G \ell} g_{ab}^{(3)} + \mathcal O(\epsilon^2),
\end{equation}
which agrees with (B.5) in \cite{Compere:2020lrt}. This result can be lifted easily to any odd dimension $d$. For the treatment of the $d=4$ case, we need to reverse-engineer the relations between $g^{(2)}_{ab}$, $\tilde g^{[4]}_{ab}$ and $R^{(2)}_{ab}$. We can show that
\begin{equation}
\begin{split}
R_{ab}[\gamma] &= R^{(0)}_{ab} + \epsilon^2 \frac{4\eta}{\ell^2} \left[ \tilde g_{ab}^{[4]} - \frac{1}{2} g_a^{(2)c}g_{bc}^{(2)} + \frac{1}{8} (g^{cd}_{(2)}g^{(2)}_{cd})g^{(0)}_{ab} \right] + \mathcal O(\epsilon^3)  , \\
R[\gamma] &= \epsilon^2 R^{(0)} + \epsilon^4 \frac{2\eta}{\ell^2} \left[ g^{cd}_{(2)} g^{(2)}_{cd} + \frac{1}{2}(g^{cd}_{(0)}g^{(2)}_{cd})^2 \right] + \mathcal O(\epsilon^5)   .
\end{split}
\end{equation}
This helps in deriving
\begin{equation}
\texttt T^{[4]}_{ab}[\gamma] = \epsilon^2\frac{\eta}{4\pi G\ell} \left( g^{(4)}_{ab} + X^{[4]}_{ab} + \frac{3}{2}\tilde g_{ab}^{[4]} \right)+\mathcal O(\epsilon^4) \label{T4_with_gtilde}
\end{equation}
where $X^{[4]}_{ab}$ is given by \eqref{X4}. The last term in \eqref{T4_with_gtilde} can be removed by means of the freedom to add a finite boundary term $L_\circ = \mathcal O(\epsilon^0)$ to $L^\epsilon_{ct}$. The key point for this procedure is the conformal anomaly. We define the boundary action
\begin{equation}
S_\circ^\epsilon = \int_{\rho=\epsilon} \bm L_\circ \equiv \frac{\eta\, \kappa}{16\pi G\ell} \int_{\rho=\epsilon} \D^d x\sqrt{|g^{(0)}|}\, \tilde a_{[d]}
\end{equation}
where $\kappa\in\mathbb R$ will be fixed herebelow. Since one can show that \cite{deHaro:2000vlm}
\begin{equation}
\tilde g_{ab}^{[d]} = \frac{1}{\sqrt{|g^{(0)}|}} \frac{2}{d} \frac{\delta}{\delta g_{(0)}^{ab}} \Big( \sqrt{|g^{(0)}|}\, \tilde a_{[d]} \Big)  ,
\label{gtilde in terms of action tilde a}
\end{equation}
we see that it is sufficient to add $\bm L_\circ$ to $\bm L_{ct}$ with $\kappa = 3/2$ to remove the $\tilde g_{ab}^{[4]}$ field from \eqref{T4_with_gtilde}, leading to the final stress-tensor for $d=4$ given by \eqref{holographic stress-energy tensor} and \eqref{X4}. The equation \eqref{gtilde in terms of action tilde a} has been explicitly checked for $d=4$ using the equations of motion of quadratic gravity summarized in the appendix \ref{New massive gravity}. We are thus back to the canonical form \eqref{holographic stress-energy tensor}, which ends our review and discussion about holographic renormalization in $d+1$ dimensions.

\section{Derivation of the infinitesimal charges}
\label{app:Surface charges}

In this appendix, we explicitly check that the expression \eqref{surface charge expression} of the infinitesimal charges satisfies \eqref{fundamental formula}. In the computation below, we omit the reference to the subleading terms $\mathcal{O}(\rho)$. We start by computing the right-hand side of \eqref{fundamental formula}. Taking \eqref{eq:action solution space} and \eqref{eq:action solution space 2} into account, the variations $\delta_\xi \sqrt{|g^{(0)}|}$ and $\delta_\xi T^{ab}_{[d]}$ are given by
\begin{align}
\delta_\xi \sqrt{|g^{(0)}|} &= \frac{1}{2}\sqrt{|g^{(0)}|} \, g^{ab}_{(0)}\delta_\xi g_{ab}^{(0)} = \sqrt{|g^{(0)}|} ( D_a \bar\xi^a - d\, \sigma ), \label{eq:deltaG0} \\
\delta_\xi T^{ab}_{[d]} &= \mathcal L_{\bar\xi} T^{ab}_{[d]} + (d+2) \sigma T^{ab}_{[d]} + A^{ab}_{[d]}[\sigma]. \label{eq:deltaTab}
\end{align}
Recalling that $D_a T^{ab}_{[d]} =0$ on-shell and writing $g^{ab}_{(0)} T_{ab}^{[d]} = \mathcal{T}^{[d]}$, we get
\begin{align}
-\delta \Theta^\rho_{{ren}}[g; \mathcal \delta_\xi g] =&  \delta\left(\sqrt{|g^{(0)}|} T^{ab}_{[d]}\right)D_a \bar\xi_b +  \sqrt{|g^{(0)}|} T^{ab}_{[d]} \delta \left(D_a \bar\xi_b\right) - \delta \left( \sqrt{|g^{(0)}|} \mathcal{T}^{[d]} \sigma  \right),\\
\delta_\xi \Theta_{{ren}}^\rho [g ; \delta g] =& - \frac{1}{2} \sqrt{|g^{(0)}|} \left( D_c \bar\xi^c \, T^{ab}_{[d]} + \mathcal L_{\bar\xi} T^{ab}_{[d]}\right)\delta g_{ab}^{(0)} -  \sqrt{|g^{(0)}|} T^{ab}_{[d]}\delta (D_a\bar\xi_b) \nonumber \\
& - \frac{1}{2} \sqrt{|g^{(0)}|} A^{ab}_{[d]}[\sigma] \delta g_{ab} + \sqrt{|g^{(0)}|} \mathcal{T}^{[d]} \delta \sigma.
\end{align} Putting all together, we obtain
\begin{equation}
\begin{split}
\omega^\rho_{ren} [g; \delta_\xi g, \delta g] =&  \delta\left(\sqrt{|g^{(0)}|} T^{ab}_{[d]}\right)D_a \bar\xi_b  - \frac{1}{2} \sqrt{|g^{(0)}|} \left( D_c \bar\xi^c\, T^{ab}_{[d]} + \mathcal L_{\bar\xi} T^{ab}_{[d]}\right)\delta g_{ab}^{(0)} \\
& - \delta \left( \sqrt{|g^{(0)}|} \mathcal{T}^{[d]}  \right) \sigma  - \frac{1}{2} \sqrt{|g^{(0)}|} A^{ab}_{[d]}[\sigma] \delta g_{ab}.
\end{split}
\end{equation}
Let us now compute the left-hand side of \eqref{fundamental formula}. We define the universal part
\begin{equation}
K^{\rho a}_{\bar \xi}[g;\delta g] =  \delta \left( \sqrt{|g^{(0)}|} g_{(0)}^{ac} T^{[d]}_{bc} \right) \bar\xi^b - \frac{1}{2} \sqrt{|g^{(0)}|} \, \bar\xi^a \, T^{bc}_{[d]} \delta g_{bc}^{(0)} ,
\end{equation} $W^{[d]\rho a}_\sigma = 0$ for odd $d$, and  
\begin{equation}
W^{[2]\rho a}_\sigma[g;\delta g] = -\frac{\ell}{16 \pi G} \left[ \sqrt{|g^{(0)}|} \partial_b \sigma \delta g^{ab}_{(0)} + 2 \delta \sqrt{|g^{(0)}|} \partial^a \sigma \right] - \ell \sigma  \Theta^a_{EH} [g^{(0)}; \delta g^{(0)}] ,
\end{equation}
\begin{align}
W^{[4]\rho a}_\sigma[g;\delta g] =& \eta \frac{\ell^3}{16 \pi G} \left[ \frac{1}{6} \sqrt{|g^{(0)}|} R_{(0)} D_b \sigma \delta g^{ab}_{(0)} + \frac{1}{3} R_{(0)} D^a \sigma \delta \sqrt{|g^{(0)}|}   \right. \nonumber \\
& \left. \quad\quad\quad - \frac{1}{2} R^{ac}_{(0)} \delta \sqrt{|g^{(0)}|} D_c \sigma + \frac{1}{4} \sqrt{|g^{(0)}|} R_{cb}^{(0)} D^a \sigma \delta g^{bc}_{(0)} - \frac{1}{2} \sqrt{|g^{(0)}|} R^{(0)a}_c D_b \sigma \delta g^{bc}_{(0)} \right] \nonumber \\
&- \eta \frac{\ell^3}{4} \sigma \left[  \Theta^a_{QCG(1)}[g^{(0)}; \delta g^{(0)}]  - \frac{1}{3}\Theta^a_{QCG(2)}[g^{(0)}; \delta g^{(0)}]  \right] 
\end{align} where $\Theta^a_{EH}$, $\Theta^a_{QCG(1)}$ and $\Theta^a_{QCG(2)}$ are defined in equation \eqref{NMG potentials}. Putting all together, we write
\begin{equation}
k_{ren, \xi}^{\rho a}[g;\delta g] = K^{\rho a}_{\bar \xi} [g;\delta g]+ W^{(d)\rho a}_\sigma [g;\delta g]. 
\label{krhoa}
\end{equation} We have
\begin{equation}
\begin{split}
\partial_a K^{\rho a}_{\bar \xi}[g ; \delta g] &=  \delta \left(\sqrt{|g^{(0)}|}T^{ab}_{[d]}\right)D_a  \bar\xi_b + \sqrt{|g^{(0)}|} T^{ab}_{[d]} \delta g_{bc}^{(0)} D_a \bar\xi^c \\
&\quad - \frac{1}{2} \sqrt{|g^{(0)}|} D_a \bar\xi^a T^{bc}_{[d]} \delta g_{bc}^{(0)} - \frac{1}{2} \sqrt{|g^{(0)}|} \bar\xi^a D_a  T^{bc}_{[d]}\delta g_{bc}^{(0)}.
\end{split}
\end{equation} Furthermore, 
\begin{equation}
\partial_a W^{[d]\rho a}_\sigma [g;\delta g] =  - \delta \left( \sqrt{|g^{(0)}|} \mathcal{T}^{[d]}  \right) \sigma  - \frac{1}{2} \sqrt{|g^{(0)}|} A^{ab}_{[d]}[\sigma] \delta g_{ab}.
\end{equation}
Using \eqref{trace even}, \eqref{trace 3d}, \eqref{trace 4d} and $\mathcal L_{\bar\xi} (T^{ab}_{[d]}) = \bar\xi^c D_c T^{ab}_{[d]} - 2 T_{[d]}^{c(a}D_c \bar\xi^{b)}$, we have
\begin{equation}
\begin{split}
&\partial_a k^{\rho a}_{ren ,\xi} [g; \delta g] - \omega^\rho_{ren} [g; \delta_\xi g , \delta g]\\
&\qquad\qquad = \frac{1}{2}\sqrt{|g^{(0)}|} \left(\mathcal L_{\bar\xi} T^{ab}_{[d]}\delta g_{ab}^{(0)} + 2 T^{ab}_{[d]}\delta g_{bc}^{(0)}D_a  \bar\xi^c - \bar\xi^c D_c T^{ab}_{[d]} \delta g_{ab}^{(0)}\right) = 0,
\end{split}
\end{equation}
which finishes the verification of \eqref{fundamental formula}. Integrating \eqref{krhoa} on a codimension $2$ section of $\mathscr{I}$ ($t = \text{constant}$), as in \eqref{integration of krhoa}, gives \eqref{surface charge expression}.

\section{Charge algebra and cocycle condition}
\label{Derivation of the charge algebra}

In this appendix, we provide some details of the computations leading to the charge algebra \eqref{charge algebra} and the 2-cocycle condition \eqref{2 cocycle condition}. We work at the level of the currents instead of the charges, which allows us to keep track of the boundary terms in the computation. Rewriting \eqref{total charge} and \eqref{split of the charge} in terms of the boundary currents, we have
\begin{equation}
\ndelta J^a_\xi [g] = \delta J^a_{\xi} [g] + \Xi^a_\xi [g;\delta g]  
\end{equation} where
\begin{equation}
\begin{split}
J^a_\xi [g] &=  \sqrt{|g^{(0)}|} g^{ac}_{(0)} T^{[d]}_{bc} \bar{\xi}^b , \\
\Xi_\xi^a  [g; \delta g] &=  - \frac{1}{2} \sqrt{|g^{(0)}|} \bar{\xi}^a T^{bc}_{[d]} \delta g_{bc}^{(0)} + W^{[d]a}_\sigma[g; \delta g]  - J^a_{\delta \xi} [g] .
\end{split}
\label{currents expressions}
\end{equation} Integrating these currents on a $(d-1)$-sphere at infinity $S_\infty$ gives the infinitesimal charge expressions \eqref{total charge} and \eqref{split of the charge}:
\begin{equation}
\begin{split}
H_\xi [g] &= \int_{S_\infty} (\D^{d-1} x)~ J^t_\xi [g], \\
\Xi_\xi  [g; \delta g] &= \int_{S_\infty} (\D^{d-1} x)~ \Xi_\xi^t [g; \delta g] , \\ 
W^{[d]}_\sigma[g; \delta g] &= \int_{S_\infty} (\D^{d-1} x)~ W^{[d]t}_\sigma [g; \delta g].
\end{split}
\end{equation} 
Let us start from the left-hand side of \eqref{charge algebra}. At the level of the currents, a computation shows that
\begin{equation}
\begin{split}
&\delta_{\xi_2} J^a_{\xi_1}[g] - \frac{1}{2} \sqrt{|g^{(0)}|} \bar{\xi}^a_2 T^{bc}_{[d]} \delta_{\xi_1} g_{bc}^{(0)}  - J^a_{\delta_{\xi_1} \xi_2} [g] \\
&\quad = J^a_{[\xi_1, \xi_2]_\star}[g]+ \sqrt{|g^{(0)}|} \left( \sigma_1 \bar{\xi}^a_2 \mathcal{T}^{[d]} + g^{ac}_{(0)} A_{bc}^{[d]} \bar{\xi}^b_1 \right) + \partial_b L_{\xi_1, \xi_2}^{[ab]}[g] 
\end{split}
\end{equation} where $L^{[ab]}_{\xi_1,\xi_2}[g] = 2 \sqrt{|g^{(0)}|} T_{cd}^{[d]}g^{d[b}_{(0)} \bar{\xi}^{a]}_2 \bar{\xi}^c_1$ is a total derivative term that will not contribute when integrating on the $(d-1)$-sphere. We already notice from this expression that, for odd $d$, the algebra with the modified Lie bracket closes without extension. Indeed, $W^{[2k+1]a}_\sigma[g; \delta g] = 0$, $A_{bc}^{[2k+1]}[\sigma] = 0$ and $\mathcal T^{[2k+1]} = 0$ for any $k \in\mathbb N_0$.  

Let us now take the $W^{[d]a}_\sigma[g; \delta g]$ term appearing in $\Xi_\xi^a  [g; \delta g]$ into account (see equation \eqref{currents expressions}). After a lengthy computation, we find
\begin{equation}
\delta_{\xi_2} J^a_{\xi_1}[g] + \Xi^a_{\xi_2} [g; \delta_{\xi_1} g] = J^a_{[\xi_1, \xi_2]_\star}[g] + K_{\xi_1, \xi_2}^{[d]a}[g] + \partial_b L^{[ab]}_{\xi_1, \xi_2} [g] + \partial_b M^{[ab]}_{\xi_1, \xi_2} [g].
\end{equation} As already announced, for $d= 2k+1$ dimensions ($k \in\mathbb N_0$), we have $M^{[ab]}_{\xi_1,\xi_2}[g] = 0$ and $K_{\xi_1, \xi_2}^{[2k+1]a}[g]=0$. Now, for $d= 2$, the total derivative term takes the form
\begin{equation}
M^{[ab]}_{\xi_1, \xi_2} [g] = \frac{\ell}{8\pi G} \sqrt{|g^{(0)}|} \left( 2 \bar{\xi}^{[a}_1 D^{b]} \sigma_2 + D^{[a} \bar{\xi}^{b]}_1 \sigma_2 \right)
\end{equation} while the field-dependent 2-cocycle is given by 
\begin{equation}
\begin{split}
K_{\xi_1, \xi_2}^{[2]a}[g] &= {}_{(1)} K_{\xi_1, \xi_2}^{[2]a} [g]  + {}_{(2)} K_{\xi_1, \xi_2}^{[2]a} [g] ,\\
{}_{(1)} K_{\xi_1, \xi_2}^{[2]a} [g] &= \frac{\ell }{8 \pi G} \sqrt{|g^{(0)}|} \left( \sigma_1 D^a \sigma_2 - \sigma_2 D^a \sigma_1  \right) , \\
{}_{(2)} K_{\xi_1, \xi_2}^{[2]a} [g] &= \frac{\ell }{16 \pi G} \sqrt{|g^{(0)}|}~ R_{(0)} \left( \sigma_1 \bar{\xi}^a_2 - \sigma_2 \bar{\xi}^a_1 \right).
\end{split} \label{2 cocycle d=3}
\end{equation} For $d=4$, the total derivative term takes the form 
\begin{equation}
\begin{split}
M^{[ab]}_{\xi_1, \xi_2} [g] = \frac{\eta\,\ell^3}{16\pi G}\sqrt{|g^{(0)}|} & \left[ \frac{1}{6} D^{[a}(\sigma_2 R^{(0)}) \bar\xi_1^{b]} - \frac{1}{3} \sigma_2 R^{(0)} D^{[a}\bar\xi^{b]}_1 + \sigma_2 R^{c[a}_{(0)} D_c \bar\xi_1^{b]} \right. \\
&\,\ \left. - \sigma_2 D^{[a} {R^{b]}_{(0)c}} \bar\xi^c_1 + \frac{1}{2} R^{(0)} D^{[a} \sigma_2 \bar\xi_1^{b]} + R^{c[a}_{(0)} ( D^{b]}\sigma_2 \bar\xi^1_c -D_c\sigma_2\bar\xi^{b]}_1 ) \right]
\end{split}
\end{equation}
while the field-dependent 2-cocycle is now given by  
\begin{equation}
\begin{split}
K_{\xi_1, \xi_2}^{[4]a}[g] &= {}_{(1)} K_{\xi_1, \xi_2}^{[4]a} [g]  + {}_{(2)} K_{\xi_1, \xi_2}^{[4]a} [g] ,\\
{}_{(1)} K_{\xi_1, \xi_2}^{[4]a} [g] &= \frac{\eta\,\ell^3}{16\pi G}\sqrt{|g^{(0)}|}  \left( R^{ab}_{(0)}-\frac{1}{2}R^{(0)}g^{ab}_{(0)}\right) \left(\sigma_1 D_b \sigma_2 - \sigma_2 D_b \sigma_1\right) , \\
{}_{(2)} K_{\xi_1, \xi_2}^{[4]a} [g] &= \frac{\eta\,\ell^3}{64\pi G}\sqrt{|g^{(0)}|} \left( R^{bc}_{(0)}R_{bc}^{(0)} - \frac{1}{3} R^2_{(0)}\right) \left( \sigma_1 \bar{\xi}^a_2 - \sigma_2 \bar{\xi}^a_1 \right).
\end{split} \label{2 cocycle d=4}
\end{equation}

Let us now prove explicitly that \eqref{2 cocycle d=3} satisfies the 2-cocycle condition \eqref{2 cocycle condition} in the $d=2$ case. Let us first consider the part (1). We have
\begin{equation}
\begin{split}
&{}_{(1)} K_{[\xi_1, \xi_2]_\star, \xi_3}^{[2]a}[g] + \delta_{\xi_3} \left( {}_{(1)} K_{\xi_1, \xi_2}^{[2]a}[g]\right) + \text{cyclic(1,2,3)} \\
&\quad = \partial_b \left( 2~ \bar\xi^{[b}_3 {}_{(1)} K^{a]}_{\xi_1, \xi_2} [g]  \right) + \frac{\ell }{8 \pi G} \sqrt{|g^{(0)}|} \bar\xi^a_3 \left(\sigma_1 D^c D_c \sigma_2 - \sigma_2 D^c D_c \sigma_1 \right) + \text{cyclic(1,2,3)}.
\end{split}
\label{cocycle condition 3d1}
\end{equation} Now, the part (2) yields
\begin{equation}
\begin{split}
&{}_{(2)} K_{[\xi_1, \xi_2]_\star, \xi_3}^{[2]a} [g] + \delta_{\xi_3} \left( {}_{(2)} K_{\xi_1, \xi_2}^{[2]a} [g]\right) + \text{cyclic(1,2,3)} \\
&\quad = \partial_b \left( \bar\xi^{[b}_3 {}_{(2)} K^{a]}_{\xi_1, \xi_2}[g]  \right) - \frac{\ell }{8 \pi G} \sqrt{|g^{(0)}|}\bar\xi^a_3 \left(\sigma_1 D^c D_c \sigma_2 - \sigma_2 D^c D_c \sigma_1 \right) + \text{cyclic(1,2,3)}.
\end{split}
\label{cocycle condition 3d2}
\end{equation} Putting \eqref{cocycle condition 3d1} and \eqref{cocycle condition 3d2} together, we finally obtain
\begin{equation}
\begin{split}
&K_{[\xi_1, \xi_2]_\star, \xi_3}^{[2]a}[g] + \delta_{\xi_3} K_{\xi_1, \xi_2}^{[2]a}[g] + \text{cyclic(1,2,3)} \\
&\quad = \partial_b \left( 2 \bar\xi^{[b}_3 {}_{(1)} K^{a]}_{\xi_1, \xi_2} [g]  + \bar\xi^{[b}_3 {}_{(2)} K^{a]}_{\xi_1, \xi_2} [g]  \right) + \text{cyclic(1,2,3)} 
\end{split}
\end{equation} where the right-hand side is a total derivative term that will disappear when integrating on the $(d-1)$-sphere. The $2$-cocycle condition can be checked on the expression \eqref{2 cocycle d=4} in the $d=4$ case in a similar way.


\section{Useful formulae from quadratic curvature gravity}
\label{New massive gravity}

In this appendix, we review some material about the quadratic curvature gravity theory \cite{Stelle:1977ry , Salvio:2018crh} (also coined as new massive gravity theory \cite{Alkac:2012bz}). Indeed, the patterns of this theory appear naturally when considering the Weyl charges in the $d = 4$ case. 

The quadratic gravity Lagrangian in $d$ dimensions contains generally three pieces
\begin{equation}
\bm L_{QCG}[g] = \bm L_{EH}[g]  + \beta_1 \, \bm L_{QCG(1)}[g]  + \beta_2 \, \bm L_{QCG(2)}[g] 
\end{equation} where $g$ denotes an arbitrary metric tensor in $d$ dimensions, $\beta_1,\beta_2$ are real constants and
\begin{equation}
\begin{split}
\bm L_{EH}[g]  &=\frac{\sqrt{|g|}}{16 \pi G} \left( R + \eta \frac{d (d-1)}{\ell^2} \right) ~\D^dx, \\
\bm L_{QCG(1)}[g]  &=\frac{\sqrt{|g|}}{16 \pi G} R_{ab} R^{ab} ~\D^dx, \qquad
\bm L_{QCG(2)}[g]  = \frac{\sqrt{|g|}}{16 \pi G} R^2  ~\D^dx.
\end{split}
\end{equation} Taking the variation of these Lagrangians, we have
\begin{equation}
\begin{split}
\delta \bm L_{EH}[g]  &= \frac{\delta \bm L_{EH}[g]}{\delta g^{ab}} \delta g^{ab} + \D \bm \Theta_{EH}[g;\delta g] ,\\
\delta \bm L_{QCG(1)}[g]  &= \frac{\delta \bm L_{QCG(1)}[g]}{\delta g^{ab}} \delta g^{ab} + \D \bm \Theta_{QCG(1)}[g;\delta g], \\
\delta \bm L_{QCG(2)}[g]  &= \frac{\delta \bm L_{QCG(2)}[g]}{\delta g^{ab}} \delta g^{ab} + \D \bm \Theta_{QCG(2)}[g;\delta g]  
\end{split}
\end{equation} where
\begin{align}
 \frac{\delta \bm L_{EH}[g] }{\delta g^{ab}} &= \frac{\sqrt{|g|}}{16 \pi G}\left[ R_{ab} - \frac{1}{2} g_{ab} - \eta \frac{d(d-1)}{2 \ell^2} g_{ab} \right] ~\D^d x ,\nonumber \\
\frac{\delta \bm L_{QCG(1)}[g] }{\delta g^{ab}} &= \frac{\sqrt{|g|}}{16 \pi G} \left[ 2 R_{acbd} R^{cd} - D_a D_b R + D^c D_c R_{ab} + \frac{1}{2} g_{ab} \left( D^c D_c R - R_{cd} R^{cd} \right) \right] ~\D^dx, \nonumber \\
\frac{\delta \bm L_{QCG(2)}[g] }{\delta g^{ab}} &=\frac{\sqrt{|g|}}{16 \pi G} \left[ 2 R R_{ab} - 2 D_a D_b R + g_{ab} \left( 2 D^c D_c R - \frac{1}{2} R^2 \right) \right] ~\D^dx 
\end{align} and
\begin{align}
\bm \Theta_{EH}[g;\delta g] &= \frac{\sqrt{|g|}}{16 \pi G} \left[ D_b (\delta g)^{ab} - D^a (\delta g)^c_c \right] ~ (\D^{d-1} x)_a ,\nonumber  \\
\bm \Theta_{QCG(1)}[g;\delta g] &= \frac{\sqrt{|g|}}{16 \pi G}  \left[ 2 R^{bc} \delta \Gamma^a_{bc} - 2 R^{ab} \delta \Gamma^c_{bc} + D^a R \delta \ln \sqrt{|g|} + 2 D_c {R^a}_b \delta g^{bc} - D^a R_{bc} \delta g^{bc} \right] ~ (\D^{d-1} x)_a, \nonumber \\
\bm \Theta_{QCG(2)}[g;\delta g] &=2R \bm \Theta_{EH}[g;\delta g] +\frac{\sqrt{|g|}}{16 \pi G} \left[ 4 D^a R \delta \ln \sqrt{|g|} + 2 D_b R \delta g^{ab} \right] ~ (\D^{d-1} x)_a . \label{NMG potentials}
\end{align}

\providecommand{\href}[2]{#2}\begingroup\raggedright\endgroup

\end{document}